\documentclass[twocolumn,twocolappendix]{aastex631}
\usepackage{apjfonts} 
\usepackage{makecell}
\usepackage{graphicx}	% Including figure files
\usepackage{amsmath}	% Advanced maths commands
\usepackage{xspace}  % enable xspace command
\usepackage{amsmath}
\usepackage{CJKutf8}
\usepackage{multirow}  % multi-rows
\usepackage{tabularx}
\usepackage{makecell}
\usepackage{supertabular,booktabs}
\renewcommand{\toprule}{\hline\hline}
\renewcommand{\midrule}{\hline}
\renewcommand{\bottomrule}{\hline}

\newcommand{\mASIAA}{Institute of Astronomy and Astrophysics, Academia Sinica, No. 1, Sec. 4, Roosevelt Road, Taipei 10617, Taiwan}
\newcommand{\ASIAA}{\affiliation{\mASIAA}}

\newcommand{\mCASS}{Center for Astrophysics and Space Sciences, Department of Physics, University of California, San Diego, 9500 Gilman Drive, La Jolla, CA 92093, USA}
\newcommand{\CASS}{\affiliation{\mCASS}}
\newcommand{\mUCSD}{Department of Astronomy \& Astrophysics, University of California, San Diego, 9500 Gilman Drive, La Jolla, CA 92093, USA}
\newcommand{\UCSD}{\affiliation{\mUCSD}}
\newcommand{\mCFA}{Center for Astrophysics $|$ Harvard \& Smithsonian, 60 Garden Street, Cambridge, MA 02138, USA}
\newcommand{\CFA}{\affiliation{\mCFA}}
\newcommand{\mMaryland}{Department of Astronomy, University of Maryland, College Park, MD 20742, USA}
\newcommand{\Maryland}{\affiliation{\mMaryland}}

\newcommand{\mMPIA}{Max-Planck-Institut f\"ur Astronomie, K\"onigstuhl 17, D-69117 Heidelberg, Germany}
\newcommand{\MPIA}{\affiliation{\mMPIA}}
\newcommand{\mOSU}{Department of Astronomy, The Ohio State University, 4055 McPherson Laboratory, 140 West 18th Avenue, Columbus, OH 43210, USA}
\newcommand{\OSU}{\affiliation{\mOSU}}

\newcommand{\mCCAPP}{Center for Cosmology and Astroparticle Physics, 191 West Woodruff Avenue, Columbus, OH 43210, USA}
\newcommand{\mPrinceton}{Department of Astrophysical Sciences, Princeton University, 4 Ivy Lane, Princeton, NJ 08544, USA}
\newcommand{\Princeton}{\affiliation{\mPrinceton}}
\newcommand{\CCAPP}{\affiliation{\mCCAPP}}

\newcommand{\mUGENT}{Sterrenkundig Observatorium, Universiteit Gent, Krijgslaan 281 S9, B-9000 Gent, Belgium}
\newcommand{\UGENT}{\affiliation{\mUGENT}}
\newcommand{\mMcMaster}{Department of Physics and Astronomy, McMaster University, 1280 Main Street West, Hamilton, ON L8S 4M1, Canada}
\newcommand{\McMaster}{\affiliation{\mMcMaster}}
\newcommand{\mCITA}{Canadian Institute for Theoretical Astrophysics (CITA), University of Toronto, 60 St George Street, Toronto, ON M5S 3H8, Canada}
\newcommand{\CITA}{\affiliation{\mCITA}}
\newcommand{\mOxford}{Sub-department of Astrophysics, Department of Physics, University of Oxford, Keble Road, Oxford OX1 3RH, UK}
\newcommand{\Oxford}{\affiliation{\mOxford}}

\newcommand{\SigmaMassUnit}{\ensuremath{\rm M_\odot~pc^{-2}}\xspace}

\newcommand{\SigmasfrUnit}{\ensuremath{{\rm M_\odot\,yr}^{-1}\,{\rm kpc}^{-2}}\xspace}
\newcommand{\ICOUnit}{{\rm K\,km\,s\textsuperscript{-1}}\xspace}
\newcommand{\SEDUnit}{{\rm MJy\,sr\textsuperscript{-1}}\xspace}
\newcommand{\acoUnit}{\ensuremath{{\rm M_\odot\,pc^{-2}\,(K\,km\,s^{-1})^{-1}}}\xspace}
\newcommand{\XcoUnit}{\ensuremath{\rm cm^{-2}\,(K\,km\,s^{-1})^{-1}}\xspace}

\newcommand{\rms}{\ensuremath{\Delta_\mathrm{rms}}\xspace}
\newcommand{\metal}{12+\logt({\rm O/H})\xspace}
\newcommand{\CO}{{\rm CO}\xspace}

\newcommand{\HI}{\textsc{H\,i}\xspace}
\newcommand{\HII}{\textsc{H\,ii}\xspace}

\newcommand{\R}[1]{{\rm R}\textsubscript{#1}\xspace}
\newcommand{\Sigmad}{\ensuremath{\Sigma_\mathrm{dust}}\xspace}
\newcommand{\Sigmastar}{\ensuremath{\Sigma_\star}\xspace}
\newcommand{\Sigmasfr}{\ensuremath{\Sigma_\mathrm{SFR}}\xspace}
\newcommand{\Sigmagas}{\ensuremath{\Sigma_\mathrm{gas}}\xspace}

\newcommand{\Sigmaatom}{\ensuremath{\Sigma_\mathrm{atom}}\xspace}
\newcommand{\Sigmamol}{\ensuremath{\Sigma_\mathrm{mol}}\xspace}
\newcommand{\Sigmatot}{\ensuremath{\Sigma_\mathrm{Total}}\xspace}

\newcommand{\aco}{\ensuremath{\alpha_\CO}\xspace}
\newcommand{\acoMW}{\ensuremath{\alpha_\CO^{\rm MW}}\xspace}

\newcommand{\Ubar}{\ensuremath{\overline{U}}\xspace}
\newcommand{\Umin}{\ensuremath{U_\mathrm{min}}\xspace}

\newcommand{\shortminus}{\scalebox{0.75}[1.0]{\ensuremath{-}}}
\newcommand{\coone}{\textup{CO}\,\ensuremath{(1{\shortminus}0)}\xspace}
\newcommand{\cotwo}{\textup{CO}\,\ensuremath{(2{\shortminus}1)}\xspace}
\newcommand{\acoone}{\ensuremath{\alpha_{\coone}}\xspace}
\newcommand{\acotwo}{\ensuremath{\alpha_{\cotwo}}\xspace}
\newcommand{\wcoone}{\ensuremath{W_{\coone}}\xspace}
\newcommand{\wcotwo}{\ensuremath{W_{\cotwo}}\xspace}

\newcommand{\nsample}{37\xspace}
\newcommand{\SigmastarT}{\ensuremath{\Sigma_{\rm\star,T}}\xspace}
\newcommand{\acoT}{\ensuremath{\alpha_{\rm CO,gal,T}}\xspace}
\renewcommand{\metal}{\ensuremath{12+\log({\rm O/H})}\xspace}
\usepackage{bm}

\shorttitle{Resolved CO-to-H$_2$ Conversion Factor in 37 Galaxies}
\shortauthors{Chiang et al.}

\begin{document}
\title{Resolved Measurements of the CO-to-H$_2$ Conversion Factor in 37 Nearby Galaxies}

\correspondingauthor{I-Da Chiang}
\email{idchiang@asiaa.sinica.edu.tw}

\author[0000-0003-2551-7148]{I-Da Chiang \begin{CJK*}{UTF8}{bkai}(江宜達)\end{CJK*}}
\ASIAA
\author[0000-0002-4378-8534]{Karin M. Sandstrom}
\UCSD
\author[0000-0002-5235-5589]{J\'er\'emy Chastenet}\UGENT
\author[0000-0002-5480-5686]{Alberto D. Bolatto}
\Maryland
\author[0000-0001-9605-780X]{Eric W. Koch}\CFA
\author[0000-0002-2545-1700]{Adam K. Leroy}\OSU\CCAPP
\author[0000-0003-0378-4667]{Jiayi Sun \begin{CJK*}{UTF8}{gbsn}(孙嘉懿)\end{CJK*}}\McMaster\CITA\Princeton
\author[0000-0003-4209-1599]{Yu-Hsuan Teng}\CASS
\author[0000-0002-0012-2142]{Thomas G. Williams}\MPIA\Oxford

\begin{abstract}
We measure the CO-to-H$_2$ conversion factor ($\alpha_\mathrm{CO}$) in 37 galaxies at 2~kpc resolution, using dust surface density inferred from far-infrared emission as a tracer of the gas surface density and assuming a constant dust-to-metals ratio. In total, we have $\sim790$ and $\sim610$ independent measurements of $\alpha_\mathrm{CO}$ for CO (2-1) and (1-0), respectively. The mean values for $\alpha_\mathrm{CO~(2-1)}$ and $\alpha_\mathrm{CO~(1-0)}$ are $9.3^{+4.6}_{-5.4}$ and $4.2^{+1.9}_{-2.0}~M_\odot~pc^{-2}~(K~km~s^{-1})^{-1}$, respectively. The CO-intensity-weighted mean for $\alpha_\mathrm{CO~(2-1)}$ is 5.69, and 3.33 for $\alpha_\mathrm{CO~(1-0)}$. We examine how $\alpha_\mathrm{CO}$ scales with several physical quantities, e.g.\ star-formation rate (SFR), stellar mass, and dust-mass-weighted average interstellar radiation field strength ($\overline{U}$).
Among them, $\overline{U}$, $\Sigma_{\rm SFR}$, and integrated CO intensity ($W_\mathrm{CO}$) have the strongest anti-correlation with spatially resolved $\alpha_\mathrm{CO}$. We provide linear regression results to \aco for all quantities tested. At galaxy integrated scales, we observe significant correlations between $\alpha_\mathrm{CO}$ and $W_\mathrm{CO}$, metallicity, $\overline{U}$, and $\Sigma_{\rm SFR}$. We also find that the normalized $\alpha_\mathrm{CO}$ decreases with stellar mass surface density ($\Sigma_\star$) in the high surface density regions ($\Sigma_\star\geq100~{\rm M_\odot~pc^{-2}}$), following the power-law relations $\alpha_\mathrm{CO~(2-1)}\propto\Sigma_\star^{-0.5}$ and $\alpha_\mathrm{CO~(1-0)}\propto\Sigma_\star^{-0.2}$. The power-law index is insensitive to the assumed dust-to-metals ratio. We interpret the decrease in $\alpha_\mathrm{CO}$ with increasing $\Sigma_\star$ as a result of higher velocity dispersion compared to isolated, self-gravitating clouds due to the additional gravitational force from stellar sources, which leads to the reduction of $\alpha_\mathrm{CO}$. The decrease in $\alpha_\mathrm{CO}$ at high $\Sigma_\star$ is important for accurately assessing molecular gas content and star-formation efficiency in the centers of galaxies, which bridge ``MW-like'' to "starburst-like" conversion factors.
\end{abstract}


\section{Introduction}\label{sec:intro}
Star formation is fueled by the molecular gas in the interstellar medium (ISM). Thus, observing the molecular ISM is essential for studies of star formation and galaxy evolution. Unfortunately, the most abundant molecule in the ISM, H$_2$, is not directly observable in many cases in the the cold molecular ISM due to its high transition energies ($h\nu/k_\mathrm{B}\sim510~\mathrm{K}$) for the lowest rotational levels.
As a result, low-$J$ CO emission lines are the most widely used tracer for the molecular ISM, as the second most abundant molecule  with strong millimeter rotational lines that can be excited at typical temperatures in molecular clouds.
The standard practice is to use a CO-to-H$_2$ conversion factor \aco, as follows
\begin{equation}\label{eq:aco}
    \Sigmamol = \left\{
    \begin{array}{l}
        \acoone \wcoone,~\mathrm{for}~\coone\\
        \acotwo \wcotwo,~\mathrm{for}~\cotwo
    \end{array}\right.\,.    
\end{equation}
where \Sigmamol [$\rm M_\odot~pc^{-2}$] is the surface density of molecular ISM (including mass of Helium), \aco [$\rm M_\odot~pc^{-2}({\rm K~km~s^{-1}})^{-1}$] is the ``CO-to-H$_2$ conversion factor'', and $W_{\rm CO}$ [${\rm K~km~s^{-1}}$] is the integrated intensity of the CO emission at rest frame frequency. The conventional \aco in the Milky Way (MW) is $\acoMW=4.35$ \acoUnit for \coone\footnote{In column density units, the standard MW conversion factor is $X_{\coone} = 2\times 10^{20}\,\XcoUnit$, where only H$_2$ is considered}. In mass surface density units with a factor of 1.36 for Helium included, this is equivalent to $\acoone = 4.35$ \acoUnit. \citep{SOLOMON87_1987ApJ...319..730S,STRONG96_1996AA...308L..21S,ABDO10_2010ApJ...710..133A}.
The ``(1\shortminus 0)''  and ``(2\shortminus 1)'' symbols represent the CO rotational transition which \aco is derived for and $W_\mathrm{CO}$ is measured, i.e.\ \coone stands for the CO~$J=1\to 0$ rotational transition at $\sim 115~\mathrm{GHz}$ ($\lambda\sim2.6~\mathrm{mm}$) and \cotwo stands for the CO~$J=2\to 1$ rotational transition at $\sim 230~\mathrm{GHz}$ ($\lambda\sim1.3~\mathrm{mm}$), respectively. In this work, we focus on the $^{12}$C$^{16}$O isotopologue and use CO for $^{12}$C$^{16}$O for simplicity.

\coone had been the most frequently measured CO transition, and \acoone is thus the fiducial case for \aco in the literature.
Meanwhile, CO emission from the next highest rotational level, i.e.\ \cotwo, has become more and more common with modern instruments, e.g.\ ALMA, throughout the last decade. Directly deriving \acotwo has attained its own importance. Thus we will present both \acoone and \acotwo in this work.

With the precise measurements of CO emission from modern instruments, our understanding of \aco has become the factor that limits our ability to precisely study the molecular ISM and star formation in nearby galaxies. Observations have shown at least two main trends in the variation of \aco. The first one is that \aco tends to increase at lower metallicity or lower dust-to-gas ratios \citep[e.g.\ up to $\sim 1~\mathrm{dex}$ higher than the MW value at $\sim 0.2$ solar metallicity,][]{ISRAEL97,LEROY11}. This enhanced \aco is often explained by the decrease in CO emission relative to the cloud mass defined by H$_2$ as shielding for CO weakened at lower metallicity \citep{Papadopoulos02_DarkGas,Grenier05_DarkGas,WOLFIRE10,Genzel12,PlanckCollaborationXIX2011,ACCURSO17,Gong20,Madden20_DarkGas,Hirashita23b_aCO}. This phenomena is often known as the ``CO-dark gas.''

The second trend is that \aco appears to be lower in the central $\sim \mathrm{kpc}$ of some galaxies \citep[it can be a factor 5--10 times lower,][]{Israel09a,Israel09b,Israel20,SANDSTROM13,Teng22,Teng23}. It is also observed to be lower in (ultra-)luminous infrared galaxies \citep[U/LIRGs,][]{DownesSolomonRadford93,DownesSolomon98,Papadopoulos12,Herrero-Illana19}. This trend towards lower \aco in galaxy centers and starbursts likely results from a combination of higher gas temperature, larger line width, and lower CO optical depth, which breaks the relationship between molecular cloud mass and line width that one would expect from isolated, virialized clouds \citep{Shetty2011b_2011MNRAS.415.3253S,BOLATTO13}. This phenomena is often referred to as the ``starburst conversion factor.'' Because galaxy centers and U/LIRGs tend to be bright in CO and thus easily observed, understanding this starburst conversion factor is important to making best use of a wide range of extragalactic observations in characterizing the star formation efficiency, gas dynamics and \HI-to-H$_2$ transition conditions.

\citet{BOLATTO13} proposed a formula for conversion factor treating the CO-dark gas and the starburst trend independently and simultaneously. This formula aimed to predict both the spatially resolved measurements from \citet[][including galaxy centers]{SANDSTROM13} and the (U)LIRGs measurements in \citet{DownesSolomon98}. The formula reads:
\begin{multline}\label{eq:acoBolatto}
    \frac{\acoone}{1~\acoUnit} = \\
    \rm 2.9\times\exp\left(\frac{0.4}{Z'}\right)\times\left\{
    \begin{array}{ll}
        \left(\Sigma_{\rm Total}^{100}\right)^{-0.5} & ,\,\Sigma_{\rm Total}^{100} \geq 1 \\
        1 & ,\,\Sigma_{\rm Total}^{100} < 1
    \end{array}\right.\,.
\end{multline}
Here $\rm Z'$ is the metallicity relative to the solar value, which traces the CO-dark gas effect; $\Sigma_{\rm Total}^{100}$ is the total surface density ($\Sigmatot = \Sigmagas + \Sigmastar$) in unit of $100~\mathrm{M_\odot~pc^{-2}}$, which is the proposed observational tracer and threshold for regions where the decrease of \aco occurs, i.e. galaxy centers and (U)LIRGs. The authors found that with a threshold at $\Sigma_{\rm Total}^{100} \geq 1$ (a threshold related to self-gravitating giant molecular clouds), the relation $\aco \propto \left(\Sigma_{\rm Total}^{100}\right)^{-0.5}$ reproduces the trend found in galaxy centers and ULIRG samples. A similar formula was also suggested by \citet{OstrikerShetty11}, where the authors suggested a power-law relationship between \aco and \Sigmagas to describe the decrease in \aco\ needed for their simulations to match observations. They showed that a relation of $\aco\propto\Sigmagas^{-0.5}$ over the surface density range $10^2$ to $10^3$~\SigmaMassUnit\ brings the observations and simulations into agreement.

To better characterize how \aco depend on local environments, spatial resolved measurements of \aco are required. To measure \aco, one needs to measure \Sigmamol independent of (single-line) CO intensity, and then divide it by the measured CO intensity. This could be achieved by several methodologies, e.g.\ using virial mass estimates \citep[see the review in][]{McKee_Ostriker_2007_review}, 
modelling multiple spectral lines \citep[e.g.][]{Cormier18,Teng22,Teng23},
converting $\gamma$-ray emission \citep[e.g.][]{ABDO10_2010ApJ...710..133A,ACKERMANN12},
and tracing gas mass with dust mass \citep{Israel97a,ISRAEL97,Leroy07,Leroy09_aCO,LEROY11,Bolatto11,PlanckCollaborationXIX2011,SANDSTROM13,Schruba17,DenBrok23}. Based on existing resources, most of the methods are not practical for a survey in tens of nearby galaxies due to the requirement in target brightness or total observing time. The most feasible methodology is to use dust as a tracer for gas mass, where dust mass can be derived from infrared (IR) data observed with \textit{Herschel} \citep{PILBRATT10}, WISE \citep{WRIGHT10}, and \textit{Spitzer}.

In this work, we measure \aco with dust as the tracer for total gas mass. We use dust masses derived from modeling the far-IR spectral energy distribution (SED) to trace the total gas mass. The key assumption we make is a constant fraction of heavy elements locked in the solid phase, i.e.\ a dust-to-metals ratio (D/M), which allows us to convert measurements of dust surface density, \HI surface density and metallicity into molecular gas mass. The assumption of approximately constant D/M is supported by dust evolution models \citep{DWEK98,HIRASHITA11,ASANO13,FELDMANN15} and kpc-scale measurements \citep{ISSA90,LEROY11,DRAINE14,Vilchez19,Chiang21} in high-metallicity ($\metal\gtrsim 8.2$) galaxy disks, matching the region of interest in this work. In simulations \citep[e.g.][]{AOYAMA20_2020MNRAS.491.3844A,Choban22}, an approximately constant D/M results from efficient dust growth in the ISM, i.e.\ the majority of the refractory elements are locked in solid grains quickly. Although there are also studies that found variations in D/M with both depletion measurements \citep{JENKINS09,JENKINS17,ROMAN-DUVAL19_2019ApJ...871..151R} and emission measurements \citep{ROMAN-DUVAL14,ROMAN-DUVAL17,CHIANG18,DEVIS19}, no widely agreed-upon prescription for the environmental dependence of D/M has been found thus far. The other reason we assume a constant D/M is that we anticipate the variation of D/M ($\leq 2$ times) to be smaller in comparison to that of \aco (up to $\sim10$ times) in normal galaxy disks.

Given the challenges in measuring spatial resolved \aco, there have been few studies with large samples of resolved measurements in galaxies. \citet{SANDSTROM13} looked at the overlap of IR from KINGFISH \citep{KENNICUTT11(KINGFISH)2011PASP..123.1347K}, CO from HERACLES \citep{LEROY09}, and \HI from THINGS \citep{WALTER08} and made measurements of \aco in 26 galaxies with $\sim$40\arcsec resolution elements, which is the resolved study with one of the largest sample size. Most other studies in the field, e.g.\ \citet{HUNT15}, \citet{ACCURSO17} and COMING \citep{Yasuda23}, focused on galaxy-scale \aco. Moreover, a survey of \aco at fixed physical scale, which allows us to evaluate the environmental dependence of \aco fairly, is also missing as previous spatially resolved studies \citep[e.g.][]{LEROY11,SCHRUBA12,SANDSTROM13} tend to perform their analysis at fixed angular resolution. 
In this work, we will measure \aco across \nsample nearby galaxies at a fixed $\sim$2~kpc resolution. This study is made possible by several surveys of resolved CO intensities in the past two decades:
the Nobeyama CO Atlas of nearby galaxies \citep[CO Atlas,][]{Kuno07_NRO_CO}, 
the HERA CO Line Extragalactic Survey \citep[HERACLES,][]{LEROY09},
the CO Multi-line Imaging of Nearby Galaxies project \citep[COMING,][]{Sorai19_COMING},
the Physics at High Angular resolution in Nearby Galaxies project \citep[PHANGS-ALMA,][]{Leroy21_PHANGS-ALMA_CO},
and recent IRAM 30m observations \citep[PI: A. Schruba; see][]{Leroy21_CO_Line_Ratios}.

This paper is presented as follows.
In \autoref{sec:calc_aco}, we explain our methodology for deriving \aco from the data. 
We describe the data sets necessary for this work and how we constrain other physical quantities from observations in \autoref{sec:data}.
We present the \aco measurements and their correlations with local and galaxy-integrated conditions in \autoref{sec:results}.
In \autoref{sec:aCO-SigmaMstar}, we investigate an observed power-law relation between \aco and \Sigmastar in the high-surface-density regions, and provide a prescription for \aco based on our findings.
We discuss the physical interpretations of our results and how it compares to the literature findings in \autoref{sec:discussions}.
Finally, we summarize our findings in \autoref{sec:summary}. 

\section{Calculating \texorpdfstring{\aco}{Conversion Factor}}\label{sec:calc_aco}
To calculate \aco, we first estimate \Sigmamol without using CO emission and an adopted conversion factor. In this study, we use a dust-based strategy to estimate \Sigmamol by assuming a value for the fraction of metals locked in solid phase, i.e.\ the dust-to-metals ratio (D/M). The D/M is defined as:
\begin{equation}\label{eq:dtm}
    {\rm D/M} = \frac{\Sigmad}{Z \times \Sigmagas},
\end{equation}
where $\Sigmagas=\Sigmaatom + \Sigmamol$ is the total neutral gas surface density, \Sigmamol is the molecular gas surface density, and the metallicity $Z$ converts \Sigmagas to a ``metal mass surface density''. By replacing \Sigmamol in the above equation with the definition of \aco in \autoref{eq:aco}, we have:
\begin{equation}\label{eq:aco_final}
    \aco = \Sigmamol~W_\mathrm{CO}^{-1} = \Big(\frac{\Sigmad}{Z\times \mathrm{(D/M)}}-\Sigmaatom\Big){W_\mathrm{CO}^{-1}},
\end{equation}
where \Sigmad, Z, \Sigmaatom, and $W_\mathrm{CO}$ are measurable quantities, thus, by assigning a value of D/M, we can then calculate \aco with our data set. The uncertainty of \aco is propagated from the uncertainties of \Sigmad, metallicity, \Sigmaatom, and $W_\mathrm{CO}$. The typical uncertainty of the pixel-by-pixel \aco measurements in this work is in the range of 0.2--0.5~dex\footnote{The lower bound of the propagated uncertainty likely results from the adopted uncertainty in metallicity.}.

Dust-based \aco measurements in the literature usually have formulae similar with \autoref{eq:dtm} but with different assumptions. For example, \citet{ISRAEL97} and \citet{Leroy09_aCO} assumed a fixed dust-to-gas ratio (D/G) in their sample galaxies to derive \aco. On the other hand, \citet[][also see \citet{LEROY11}, \citet{DenBrok23} and \citet{Yasuda23}]{SANDSTROM13} assumed that the D/G remains approximately constant in a certain spatial region, e.g.\ kpc scale or entire galaxy. With this assumption, the authors are able to derive \aco by minimizing the scatter in D/G in a group of nearby pixels, treating \aco and D/G as free parameters. This method has the advantage in not forcing the value of D/G and the disadvantage in sacrificing the spatial resolution.

Generally speaking, we could assume different D/M values in each pixel when we apply \autoref{eq:aco_final}. However, despite the previous efforts on studying the evolution of D/M \citep{DEVIS19,AOYAMA20_2020MNRAS.491.3844A,PerouxHowk20,Chiang21,Choban22,ROMAN-DUVAL22_METAL3}, we do not have a well established prescription of how D/M depends on local environments, or the prescription is not more accurate than simply assuming a constant D/M.
Studies have shown that at $\metal > 8.2$, the D/M falls in the range\footnote{For studies that only measure D/G measurements, we quote the D/M value calculated at $Z_\odot$.} between $0.4\mbox{--}0.7$,
e.g. $0.5$ \citep[][taking $F_\star=1$]{JENKINS09},
$0.72$ \citep{LEROY11},
$0.46$ \citep{REMY-RUYER14},
$0.68$ \citep{DRAINE14},
$0.7$ \citep{FELDMANN15},
$0.56$ \citep{CHIANG18},
and
$0.40\mbox{--}0.58$ \citep{Chiang21}. Meanwhile, studies have shown that in galaxy centers, \aco can vary by a factor of 10 \citep{SANDSTROM13,BOLATTO13,Israel20,Teng22,Teng23}, which is a significantly larger dynamic range than D/M.
In the following, we will take a constant $\mathrm{D/M}=0.55$ (mean of 0.4--0.7) as the fiducial case for deriving \aco.
A possible drawback by assuming a constant D/M is that D/M has been shown to decrease toward lower metallicity \citep[e.g.][]{REMY-RUYER14,HIRASHITA11,CHIANG18,DEVIS19}. Thus, the assumed D/M value that is appropriate for galaxy centers is likely too high for low-metallicity regions (usually the outer disks), resulting in an underestimation of \aco (\autoref{eq:aco_final}) in the outer disk. 
We will discuss the case of a varying D/M with a toy model in \autoref{app:d/m}. We will also discuss the possible uncertainties in \Sigmad derivation in \autoref{sec:data}.

\subsection{\texorpdfstring{\coone}{CO (1-0)} and \texorpdfstring{\cotwo}{CO (2-1)} cases}
In this multi-wavelength, multi-galaxy study, we do not always have both the lowest-$J$ CO emission lines for all the target galaxies. Studies have been using CO line ratios to convert the intensity between CO emission lines. For an in-depth discussion on low-$J$ CO line ratios, we refer the readers to \citet{Leroy21_CO_Line_Ratios}. In the literature, perhaps the most frequently used method to treat different CO line coverage is converting everything to \wcoone with a constant CO line ratio \citep[e.g.][]{SANDSTROM13,Sun20,Chiang21}. This method allows us to uniformly use \acoone for calculating \Sigmamol in the study.
Theoretically, the line ratio can vary with excitation conditions like gas temperature and linewidth. Thus, we expect $R_{21}$ to trace the local environmental conditions. Taking the line ratio for \wcoone and \wcotwo as an example:
\begin{equation}
R_{21} \equiv \frac{\wcotwo}{\wcoone},
\end{equation}
where $R_{21}$ is usually falls in the range of 0.3 to 0.9 with mean value $\sim 0.65$ for normal star-forming galaxies, and it is expected to be higher in galaxy centers \citep{LEROY09,Leroy21_CO_Line_Ratios,Leroy23_JWST-CO,denBrok21,Yajima21}.

In this work, however, we will not adopt the simple strategy as our fiducial case because most of our target galaxies have \cotwo data, that \acotwo has attained its own importance due to modern observations, and that the variation in $R_{21}$ is non-negligible. We will present four solutions of \aco in parallel, two without any conversions between \cotwo and \coone, and two with different prescriptions of the line ratio:
\begin{enumerate}
    \item \acotwo calculated with \wcotwo data only.
    \item \acoone calculated with \wcoone data only.
    \item \acoone calculated with \wcoone data, plus \wcotwo for galaxies without \wcoone data, converted with a constant $R_{21}$.
    \item \acoone calculated with \wcoone data, plus \wcotwo for galaxies without \wcoone data, converted with an environment-dependent $R_{21}$.
\end{enumerate}
For the third method, we adopt the constant $R_{21}=0.65$ from \citet{Leroy21_CO_Line_Ratios}. For the last method, we adopt the MIR-dependent formula suggested by \citet{Leroy23_JWST-CO}, namely
\begin{equation}\label{eq: R21(W4)}
    R_{21} = 0.62\left(\frac{I_\mathrm{WISE4}}{1\,\SEDUnit}\right)^{0.22}.
\end{equation}
We follow the suggestion in \citet{Leroy23_JWST-CO} and cap $R_{21}$ at $R_{21}=1$. This formula in general agrees with the finding that $R_{21}$ scales with MIR intensity or \Sigmasfr with a power-law index $\sim 0.15$--$0.2$ \citep{denBrok21,Yajima21,Leroy21_CO_Line_Ratios,Leroy23_JWST-CO}.

\section{Data}\label{sec:data}
We measure \aco in \nsample nearby galaxies in this study. To measure \aco with our dust-based methodology (\autoref{sec:calc_aco}), the data sets required are dust surface density (\Sigmad, from IR SED modelling), CO low-$J$ rotational line integrated intensity ($W_\mathrm{CO}$), atomic gas surface density (\Sigmaatom, from \HI 21~cm line emission), and metallicity ($Z$, from gas-phase oxygen abundance in \HII regions). We first select our sample galaxies from the dust catalog of $z$=0 Multiwavelength Galaxy Synthesis ($z$0MGS) \citep[][J. Chastenet et al. in preparation]{LEROY19,Chastenet21_M101}. From this large sample, we pick the 49 galaxies with both low-$J$ CO rotational line and \HI data available, including our own new \HI data sets. We design our study with a common resolution of 2~kpc, which draws a limit in sample selection at distance $\sim 20~$Mpc since the worst resolution data in our sample usually have angular resolution around $20\arcsec$. High-inclination ($>80^\circ$) targets are also excluded. We further exclude 4 galaxies\footnote{NGC~925, NGC~2403, NGC~4496A, and NGC~7793.} that satisfy all above conditions but do not satisfy the signal-to-noise ratio conditions that will be described in \autoref{sec:data:masking and completeness}. The selection yields \nsample galaxies in our sample. We list the properties of these galaxies in \autoref{tab:samples}.

\startlongtable
\begin{deluxetable*}{cccccccccccc}
\tablecaption{Galaxy Sample.\label{tab:samples}}
% \tablewidth{700pt}
% \tabletypesize{\scriptsize}
\tablehead{
Galaxy & Dist. & $i$ & P.A. & $R_{25}$\hspace{0em} & $R_e$\hspace{0em} & $\log(M_\star)$ & Type & \coone & \cotwo & 	\textsc{Hi} Ref & 12+log(O/H) Ref \\
 \\
 & [Mpc] & [\arcdeg] & [\arcdeg] & [kpc] & [kpc] & [M$_\sun$] &  &  &  &  &  \\
(1) & (2) & (3) & (4) & (5) & (6) & (7) & (8) & (9) & (10) & (11) & (12)
} 
\startdata
IC0342 & 3.5 & 31.0 & 42.0 & 10.1 & 4.4 & 10.2 & 5 & CO Atlas & \nodata & EveryTHINGS & $f.$ \\
NGC0253 & 3.7 & 75.0 & 52.5 & 14.4 & 4.7 & 10.5 & 5 & CO Atlas & PHANGS-ALMA & $c.$ & $g.$ \\
NGC0300 & 2.1 & 39.8 & 114.3 & 5.9 & 2.0 & 9.3 & 6 & \nodata & PHANGS-ALMA & $d.$ & $h.$ \\
NGC0598 & 0.9 & 55.0 & 201.0 & 8.1 & 2.4 & 9.4 & 5 & \nodata & $a.$ & $e.$ & $h.$ \\
NGC0628 & 9.8 & 8.9 & 20.7 & 14.1 & 3.9 & 10.2 & 5 & COMING & PHANGS-ALMA & THINGS & PHANGS-MUSE \\
NGC2841 & 14.1 & 74.0 & 153.0 & 14.2 & 5.4 & 10.9 & 3 & COMING & \nodata & THINGS & $g.$ \\
NGC3184 & 12.6 & 16.0 & 179.0 & 13.6 & 5.3 & 10.3 & 5 & CO Atlas & HERACLES & THINGS & $h.$ \\
NGC3198 & 13.8 & 72.0 & 215.0 & 13.0 & 5.0 & 10.0 & 5 & COMING & HERACLES & THINGS & $g.$ \\
NGC3351 & 10.0 & 45.1 & 193.2 & 10.5 & 3.1 & 10.3 & 3 & CO Atlas & PHANGS-ALMA & THINGS & PHANGS-MUSE \\
NGC3521 & 13.2 & 68.8 & 343.0 & 16.0 & 3.9 & 11.0 & 3 & CO Atlas & PHANGS-ALMA & THINGS & $g.$ \\
NGC3596 & 11.3 & 25.1 & 78.4 & 6.0 & 1.6 & 9.5 & 5 & \nodata & PHANGS-ALMA & EveryTHINGS & $g.$ \\
NGC3621 & 7.1 & 65.8 & 343.8 & 9.9 & 2.7 & 10.0 & 6 & \nodata & PHANGS-ALMA & THINGS & $h.$ \\
NGC3627 & 11.3 & 57.3 & 173.1 & 16.9 & 3.6 & 10.7 & 3 & CO Atlas & PHANGS-ALMA & THINGS & PHANGS-MUSE \\
NGC3631 & 18.0 & 32.4 & -65.6 & 9.7 & 2.9 & 10.2 & 5 & CO Atlas & $b.$ & EveryTHINGS & $g.$ \\
NGC3938 & 17.1 & 14.0 & 195.0 & 13.4 & 3.7 & 10.3 & 5 & COMING & HERACLES & HERACLES-VLA & $g.$ \\
NGC3953 & 17.1 & 61.5 & 12.5 & 15.2 & 5.3 & 10.6 & 4 & \nodata & $b.$ & EveryTHINGS & $g.$ \\
NGC4030 & 19.0 & 27.4 & 28.7 & 10.5 & 2.1 & 10.6 & 4 & COMING & \nodata & EveryTHINGS & $g.$ \\
NGC4051 & 17.1 & 43.4 & -54.8 & 14.7 & 3.7 & 10.3 & 3 & CO Atlas & $b.$ & EveryTHINGS & $g.$ \\
NGC4207 & 15.8 & 64.5 & 121.9 & 3.5 & 1.4 & 9.6 & 7 & \nodata & PHANGS-ALMA & PHANGS-VLA & $g.$ \\
NGC4254 & 13.1 & 34.4 & 68.1 & 9.6 & 2.4 & 10.3 & 5 & CO Atlas & PHANGS-ALMA & HERACLES-VLA & PHANGS-MUSE \\
NGC4258 & 7.6 & 68.3 & 150.0 & 18.8 & 5.9 & 10.7 & 4 & COMING & \nodata & HALOGAS & $h.$ \\
NGC4321 & 15.2 & 38.5 & 156.2 & 13.5 & 5.5 & 10.7 & 3 & CO Atlas & PHANGS-ALMA & HERACLES-VLA & PHANGS-MUSE \\
NGC4450 & 16.8 & 48.5 & -6.3 & 13.3 & 4.3 & 10.7 & 2 & \nodata & $b.$ & EveryTHINGS & $g.$ \\
NGC4501 & 16.8 & 60.1 & -37.8 & 21.1 & 5.2 & 11.0 & 3 & CO Atlas & \nodata & EveryTHINGS & $g.$ \\
NGC4536 & 16.2 & 66.0 & 305.6 & 16.7 & 4.4 & 10.2 & 3 & CO Atlas & PHANGS-ALMA & HERACLES-VLA & $g.$ \\
NGC4569 & 15.8 & 70.0 & 18.0 & 21.0 & 5.9 & 10.8 & 2 & CO Atlas & PHANGS-ALMA & HERACLES-VLA & $g.$ \\
NGC4625 & 11.8 & 47.0 & 330.0 & 2.4 & 1.2 & 9.1 & 9 & \nodata & HERACLES & HERACLES-VLA & $h.$ \\
NGC4651 & 16.8 & 50.1 & 73.8 & 9.5 & 2.4 & 10.3 & 5 & \nodata & $b.$ & EveryTHINGS & $h.$ \\
NGC4689 & 15.0 & 38.7 & 164.1 & 8.3 & 4.7 & 10.1 & 5 & CO Atlas & PHANGS-ALMA & EveryTHINGS & $g.$ \\
NGC4725 & 12.4 & 54.0 & 36.0 & 17.5 & 6.0 & 10.8 & 1 & \nodata & HERACLES & HERACLES-VLA & $g.$ \\
NGC4736 & 4.4 & 41.0 & 296.0 & 5.0 & 0.8 & 10.3 & 1 & CO Atlas & HERACLES & THINGS & $g.$ \\
NGC4941 & 15.0 & 53.4 & 202.2 & 7.3 & 3.4 & 10.1 & 1 & \nodata & PHANGS-ALMA & EveryTHINGS & $g.$ \\
NGC5055 & 9.0 & 59.0 & 102.0 & 15.5 & 4.2 & 10.7 & 4 & CO Atlas & HERACLES & THINGS & $g.$ \\
NGC5248 & 14.9 & 47.4 & 109.2 & 8.8 & 3.2 & 10.3 & 3 & CO Atlas & PHANGS-ALMA & PHANGS-VLA & $g.$ \\
NGC5457 & 6.7 & 18.0 & 39.0 & 23.4 & 13.5 & 10.3 & 5 & CO Atlas & HERACLES & THINGS & $h.$ \\
NGC6946 & 7.3 & 33.0 & 243.0 & 12.1 & 4.4 & 10.5 & 5 & CO Atlas & HERACLES & THINGS & $h.$ \\
NGC7331 & 14.7 & 76.0 & 168.0 & 19.8 & 3.7 & 11.0 & 4 & COMING & HERACLES & THINGS & $g.$ \\
\enddata
\tablecomments{(2) Distance \citep[from EDD][]{dist_galbase_EDD_TULLY09}; (3-4) inclination angle and position angle \citep{1999ApJ...523..136S,DEBLOK08,LEROY09,MUNOZMATEOS09,2009ApJ...702..277M,2013ApJ...774..126M,MAKAROV14,LANGMEIDT_2020ApJ...897..122L}; (5) isophotal radius \citep{MAKAROV14}; (6) effective radius \citep{Leroy21_PHANGS-ALMA_CO}; (7) logarithmic global stellar mass \citep{LEROY19}; (8) numerical Hubble stage T;
(9) References of CO~$J=1\to0$ observations (``\nodata'' means no CO~$J=1\to0$ data adopted in this work):
CO Atlas \citet{Kuno07_NRO_CO}; COMING \citep{Sorai19_COMING};
(10) References of CO~$J=2\to1$ observations (``\nodata'' means no CO~$J=2\to1$ data adopted in this work):
HERACLES \citet{LEROY09}; PHANGS-ALMA \citep{Leroy21_PHANGS-ALMA_CO}; 
$a.$ M33 data from \citet{GRATIER10,DRUARD14};
$b.$ New HERA data \citep[P.I.: A. Schruba; presented in][]{Leroy21_CO_Line_Ratios};
(11) References of \textsc{Hi} observations:
THINGS \citep{WALTER08}; HALOGAS \citep{Heald11_HALOGAS}; HERACLES-VLA \citep{SCHRUBA11}; PHANGS-VLA (P.I. D. Utomo; I. Chiang et al. in preparation); EveryTHINGS (P.I. K. M. Sandstrom; I. Chiang et al. in preparation);
$c.$ \citet{Puche91};
$d.$ \citet{Puche90};}
$e.$ \citet{KOCH18};
(12) References of \metal measurement:
PHANGS-MUSE \citep{Emsellem22_PHANGS-MUSE,Santoro22}; 
$f.$ private communication with K. Kreckel \citep[see][]{Chiang21};
$g.$ using the empirical formula described in \autoref{sec:data:data sets};
$h.$ data from \citet{Zurita21} compilation.
\end{deluxetable*}

\subsection{Physical Quantities and Data Sets}\label{sec:data:data sets}

% Dust
\noindent\textbf{Dust properties.} 
We obtain dust properties by fitting the dust emission spectral energy distribution (SED) to the \citet{DRAINE07} physical dust model. The details of the IR data processing and dust SED fitting are reported in \citet{Chastenet21_M101} and J. Chastenet et al. (in preparation). We briefly summarize the process below.

We obtain the dust emission SED in the IR observed by two space telescopes: the 3.4, 4.6, 12 and 22~\micron\ bands with the Wide-field Infrared Survey Explorer \citep[WISE,][]{WRIGHT10}, and the 70, 100, 160 and 250~\micron\ bands with the \textit{Herschel Space Observatory} \citep{PILBRATT10}.
The \textit{Herschel} and WISE maps are first convolved to SPIRE~250 PSF and then to a 21\arcsec\ Gaussian PSF using the SPIRE~250-to-Gauss-21\arcsec\ kernel from \citet{ANIANO11}. The 21\arcsec\ PSF is the ``moderate'' Gaussian from \citet{ANIANO11} that provides relatively high angular resolution without amplifying image artifacts. Finally, these maps are convolved to the desired resolution: a Gaussian PSF with spatial resolution corresponding to FWHM of 2~kpc.

After convolving the IR maps, we fit the dust SED with the \citet{DRAINE07} physical dust model with the dust opacity calibration derived in \citet{Chastenet21_M101}. This calibration is based on metallicity measured with ``direct'' electron-temperature-based methods, which is consistent with the strong lien calibration adopted in this work \citep[S-calibration in][]{PilyuginGrebel16}, and yields reasonable D/M. Thus, the calibration ties dust opacity, D/M and metallicity into one framework.
The complete set of data products from the fitting includes the maps of dust mass surface density (\Sigmad), interstellar radiation field (the minimum radiation field \Umin and the fraction of dust heated by the power-law radiation field $\gamma$) and the fractional dust mass in the form of polycyclic aromatic hydrocarbons ($\rm q_{PAH}$). The maximum radiation field is fixed at $U_{\rm max}=10^7$ and the power-law index for radiation field distribution is fixed at $\alpha=2$. From the fitted \Umin and $\gamma$, we can derive the dust-mass-averaged radiation field \Ubar, which is the fiducial tracer for radiation field in this work.

We also note that we assume fixed dust properties in our dust SED fitting throughout this study, which is the most frequently adopted strategy in the literature. The accuracy of \Sigmad estimates, derived by fitting the IR SED with dust emission models, can be affected by variations in the dust opacity. Interstellar dust grains are not uniform in their chemical composition, size distribution, and shape, leading to variations in their opacity \citep[e.g.][]{DRAINE07,HirashitaVoshchinnikov14,DraineHensley21}. 
In the MW, \citet{Stepnik03} found that the dust opacity increases by $\sim 3$ times from the diffuse ISM to the dense clouds. The authors argued that the increase in dust opacity is resulted from the deficit of small grains due to grain-grain coagulation.
It is challenging to measure the variation in opacity of interstellar dust since it is degenerate with the environmental dependence of \aco and D/M.
Moreover, many of the mechanisms that affects dust opacity, e.g.\ grain-grain coagulation and ice mantles, are smoothed out in kpc-scale extra-galactic studies \citep{GALLIANO18}, meaning that its variation is likely less observable than the other degenerate factors like \aco and D/M.
We note that there are extra-galactic studies that attribute all the variations in dust and gas properties to dust opacity to evaluate its variation, e.g.\ \citet{CLARK19_2019MNRAS.489.5256C} found that dust opacity change by a factor $\sim 2$ within M74 and $\sim 5$ within M83.\\

% HI
\noindent\textbf{Atomic gas surface density.} We trace the atomic gas surface density (\Sigmaatom) with the \HI 21\,cm integrated intensity ($I_\HI$), assuming the opacity is negligible \citep[e.g.,][]{WALTER08}: 
\begin{equation}\label{eq:hi}
    \frac{\Sigmaatom}{1~\SigmaMassUnit} = 
    1.36\times(1.46\times10^{-2})\times \frac{W_\HI}{1~\ICOUnit}~,
\end{equation}
where the $1.36$ factor accounts for the mass of helium.

We obtain $W_\HI$ from both literature and new data, as listed in \autoref{tab:samples}. The two new \HI surveys are the EveryTHINGS survey (P.I. K. M. Sandstrom; I. Chiang et al. in preparation) and the PHANGS-VLA survey (P.I. D. Utomo). The EveryTHINGS survey targets nearby galaxies with \textit{Herschel} photometric data but without high-resolution \HI observations, while the PHANGS-VLA survey focuses on galaxies in the Physics at High Angular resolution in Nearby Galaxies (PHANGS) project\footnote{\url{http://phangs.org/}}. 
Both surveys have their data observed with the C- and D-configurations of \textit{Karl G. Jansky Very Large Array (VLA)}\footnote{The VLA is operated by the National Radio Astronomy Observatory (NRAO), which is a facility of the National Science Foundation operated under cooperative agreement by Associated Universities, Inc.}, which yield an angular resolutions in the range of $20\arcsec$ to $30\arcsec$. Both surveys provide new high-sensitivity 21\,cm \HI observations in tens of nearby galaxies. We did not include WHISP \citep{Swaters02_WHISP} data because the galaxies that only have WHISP data have resolution coarser than 2~kpc after convolving to a circular PSF.\\

% CO
\noindent\textbf{CO low-{\boldmath $J$} rotational lines.} The integrated intensity of CO low-$J$ rotational lines traces the molecular gas surface density (\autoref{eq:aco}), and is key to this study. We use the compilation of CO mapping assembled by \citet{Leroy21_CO_Line_Ratios,Leroy23_JWST-CO} from several publicly available \coone and \cotwo data:
\begin{itemize}
    \item \coone data from the COMING survey \citep{Sorai19_COMING} and the CO Atlas \citep{Kuno07_NRO_CO}.
    \item \cotwo data from HERACLES \citep{LEROY09}, the PHANGS-ALMA survey \citep{Leroy21_PHANGS-ALMA_CO}, and a new survey observed by the IRAM-30m focused on the Virgo Cluster \citep[P.I. A. Schruba; processed in][]{Leroy21_CO_Line_Ratios}.
\end{itemize}
The source of CO data for each galaxy is listed in \autoref{tab:samples}, where \coone and \cotwo are listed separately. All these literature measurements focus on the $^{12}\mathrm{C}^{16}\mathrm{O}$ isotopologue, hereafter CO for simplicity.\\

% SFR and MStar
\noindent\textbf{Surface densities of stellar mass and SFR.} We trace the surface densities of stellar mass and SFR (\Sigmastar and \Sigmasfr, respectively) using the data and conversion formulae presented in the $z$0MGS survey \citep{LEROY19}. We utilize the $z$0MGS compilation of the background-subtracted WISE \citep{WRIGHT10} $\lambda\sim3.4$ and 22\,\micron\ (hereafter WISE1 and WISE4, respectively) data and the Galaxy Evolution Explorer \citep[GALEX,][]{MARTIN05} $\lambda\sim154~\rm nm$ (hereafter FUV) data.

We use WISE1 data to trace stellar mass surface density (\Sigmastar) with:
\begin{equation}
\frac{\Sigmastar}{1~\SigmaMassUnit} = 3.3 \times 10^2 \left(\frac{\Upsilon^{3.4}_\star}{0.5~\mathrm{M_\odot~L_\odot^{-1}}}\right) \frac{I_{\rm WISE1}}{1~\SEDUnit}~,
\end{equation}
where $\Upsilon^{3.4}_\star$ is the \Sigmastar-to-WISE1 mass-to-light ratio. The value of $\Upsilon^{3.4}_\star$ is calculated from the galaxy-by-galaxy SFR-to-WISE1 ratio, a ``specific SFR-like'' quantity, with the prescription calibrated in Appendix A of \citet{LEROY19}.

We use FUV and WISE4 data to trace the SFR surface density (\Sigmasfr) also with the conversion formula suggested by $z$0MGS \citep{LEROY19,Belfiore23}. For galaxies with both FUV and WISE4 available, we use:
\begin{multline}
\frac{\Sigmasfr}{1~\SigmasfrUnit} = \\
8.85 \times 10^{-2} \frac{I_{\rm FUV}}{1~\SEDUnit} + 3.02 \times 10^{-3} \frac{I_{\rm WISE4}}{1~\SEDUnit}~.
\end{multline}

For NGC3953 and NGC4689, where only WISE4 is available, we use:
\begin{equation}
\frac{\Sigmasfr}{1~\SigmasfrUnit} =
3.81 \times 10^{-3} \frac{I_{\rm W4}}{1~\SEDUnit}~.
\end{equation}

For both WISE and GALEX maps, we blank the regions with foreground stars identified in the $z$0MGS database. We interpolate the intensities in the blanked region with a Gaussian kernel FWHM = 22.5\arcsec (the adopted WISE and GALEX maps have FWHM = 15\arcsec) with the function \texttt{interpolate\_replace\_nans} in \texttt{astropy.convolution}. This interpolation is done on the maps before any convolution, reprojection, or unit conversion. Regarding the WISE maps, this treatment is only done for the maps used for calculating \Sigmastar and \Sigmasfr. For the WISE maps used in dust SED fitting, we refer the readers to J. Chastenet et al. (in preparation).\\

% sSFR
\noindent\textbf{Specific SFR.} With the measurements of \Sigmasfr and \Sigmastar, we calculate the specific SFR (sSFR) as:
\begin{equation}
    \frac{\rm sSFR}{\rm 1~yr^{-1}} = 1 \times 10^{-6} \times
\big(\bm{\sum}_i\frac{\Sigmasfr}{1~\SigmasfrUnit}\big)\big(\bm{\sum}_i\frac{\Sigmastar}{1~\SigmaMassUnit}\big)^{-1},
\end{equation}
where $\bm{\sum}_i$ is the summation over pixels in a galaxy. Meanwhile, we calculate the spatially resolved sSFR (rsSFR) as:
\begin{equation}
    \frac{\rm rsSFR}{\rm 1~yr^{-1}} = 1 \times 10^{-6} \times
\frac{\Sigmasfr}{1~\SigmasfrUnit}\big(\frac{\Sigmastar}{1~\SigmaMassUnit}\big)^{-1}.
\end{equation}

% Metallicity
\noindent\textbf{Metallicity.} We use the abundance of oxygen, \metal, to trace the metallicity ($Z$). We assume a fixed abundance pattern, i.e.\ a constant oxygen-to-total-metal mass ratio. The conversion from \metal to $Z$ then becomes:
\begin{equation}
    Z = 0.0134 \times 10^{\metal - 8.69},
\end{equation}
where 0.0134 and 8.69 are the adopted $Z_\odot$ and $\metal_\odot$, respectively \citep{ASPLUND09}.

We calculate \metal for each pixel as a function of the galactocentric distance by adopting the radial gradient of \metal derived from measurements in \HII regions. We use data from two surveys: the PHANGS-MUSE survey \citep{Emsellem22_PHANGS-MUSE,Groves23} and the \citet{Zurita21} compilation. We use the \citet{PilyuginGrebel16} S-calibration\footnote{\citet{PilyuginGrebel16} utilizes the $S_2=I_{\rm [S~II]}\lambda 6717 + \lambda 6731/I_{\rm H\beta}$, $N_2=I_{\rm [NII]}\lambda 6548+ \lambda 6584/I_{\rm H\beta}$, and $R_3=I_{\rm [OIII]}\lambda 4959+\lambda 5007/I_{\rm H\beta}$ line ratios to measure gas-phase \metal.} (hereafter PG16S) as the fiducial calibration for \metal. PG16S is a calibration method that shows good agreement with direct metallicity measurements \citep{CROXALL16,Kreckel19}. Since PG16S only relies on lines covered by MUSE, the \metal measurement can be expanded to the full PHANGS-MUSE dataset in our future works.

For galaxies in the PHANGS-MUSE survey, we adopt radial \metal gradients presented in \citet{Santoro22}, which are calculated with the PG16S calibration. For galaxies that only appear in the \citet{Zurita21} emission data compilation, we use the \citet{Zurita21} data to calculate the PG16S \metal and then fit the radial \metal gradient in these galaxies. We only consider galaxies that have at least 5 measurements spanning at least $0.5R_{25}$ in the \citet{Zurita21} data table. The uncertainties of the \metal gradient is not explicitly provided in either work. We will assume a 0.1~dex scatter for \metal derived from gradients as suggested in \citet{Berg20}.

For galaxies without measurements of \metal in either \citet{Zurita21} or \citet{Santoro22}, we use the two-step strategy proposed in \citet{Sun20} to estimate their \metal. First, we use a mass-metallicity relation to predict \metal at one effective radius ($R_\mathrm{e}$) in a given galaxy. Second, we extend the prediction with a radial gradient of $-0.1~\mathrm{dex}/R_\mathrm{e}$ suggested by \citet{Sanchez14}. We characterize the mass-metallicity with a function of the form:
\begin{equation}
    \metal = a + bxe^{-x},
\end{equation}
where $x=\log(M_\star/\mathrm{M}_\odot) - 11.5$ \citep[see][and references therein]{Sanchez19}. $a$ and $b$ are free parameters. We fit the function with \metal at one $\rm R_e$ from galaxies with the available measurements listed in \autoref{tab:samples}. The best-fit parameters are $a=8.56\pm 0.02$ and $b=0.010\pm0.002$, which are robust under resampling. Meanwhile, the typical statistical uncertainty in the \metal data used for fitting is $\sim 0.013$~dex. However, the root mean square error (\rms) between the best fit and data is 0.13~dex, meaning that there is still intrinsic scatter in \metal that is not explained by the mass-metallicity relation and the adopted radial gradients, e.g.\ the azimuthal variations \citep{Williams22_Metal2D}. 
We use this fitted relation to derive the radial metallicity gradient of galaxies without metallicity measurements with the $M_\star$ and $R_\mathrm{e}$ listed in \autoref{tab:samples}. We will assume an uncertainty of 0.15~dex (rounding up $0.013+0.13$~dex) for galaxies with this type of \metal measurements.

Studies have reported that the PG16S calibration could result in \metal value lower than other calibrations \citep[e.g.][]{DEVIS19}. Aligning with that, there are also studies reporting that the \metal calibrated with PG16S in Orion Nebula and other \HII regions in the solar neighborhood have values $\sim0.2$~dex lower than the solar reference value \citep[e.g.][]{Esteban22}. This effect could lead to an underestimate of Z' and thus an overestimate in \aco with our methodology. For consistency with the direct metallicity calibration used in \citet{Chastenet21_M101} for calibrating dust opacity, we will use PG16S in this work.

\subsection{Uniform Data Processing}\label{sec:data:data processing}
The analyses in this work are done at a common resolution of 2~kpc for all data.
For \HI, CO, \Sigmastar, and \Sigmasfr, we convolve them to a circular Gaussian PSF with a FWHM corresponding to 2\,kpc, using the \texttt{astropy.convolution} package \citep{ASTROPY13,Astropy18,Astropy22}. The images are then reprojected onto a grid with a pixel size of one third of the FWHM (that is, we oversample at roughly the Nyquist sampling rate) with the \texttt{astropy} affiliated package \texttt{reproject}.
The convolution and reprojection of the dust maps are done in J. Chastenet et al. (in preparation). They convolve the IR maps into the final resolution using kernels from \citet{ANIANO11}. Note that the convolution is done before SED fitting for dust properties.
The galactocentric radius and metallicity are directly calculated on the final pixel grids.
All the surface density and surface brightness quantities presented in this work have been corrected for inclination, listed in \autoref{tab:samples}.

\subsection{S/N Mask and Completeness}\label{sec:data:masking and completeness}

\noindent\textbf{S/N Mask.} For statistical quantities that only involve \aco, e.g.\ mean values and percentiles, we masked out the low signal-to-noise ratio (S/N) pixels. Specifically, we blank pixels with $\mathrm{S/N} < 1$ in $W_\mathrm{CO}$ and \Sigmamol. Note that \Sigmamol here is derived from \Sigmad (with IR photometry), metallicity and \Sigmaatom (\autoref{eq:aco_final}), thus the uncertainty of \Sigmamol is propagated from the uncertainties of these three quantities and the IR photometry.\\

\begin{figure*}
    \centering
	\includegraphics[width=0.99\textwidth]{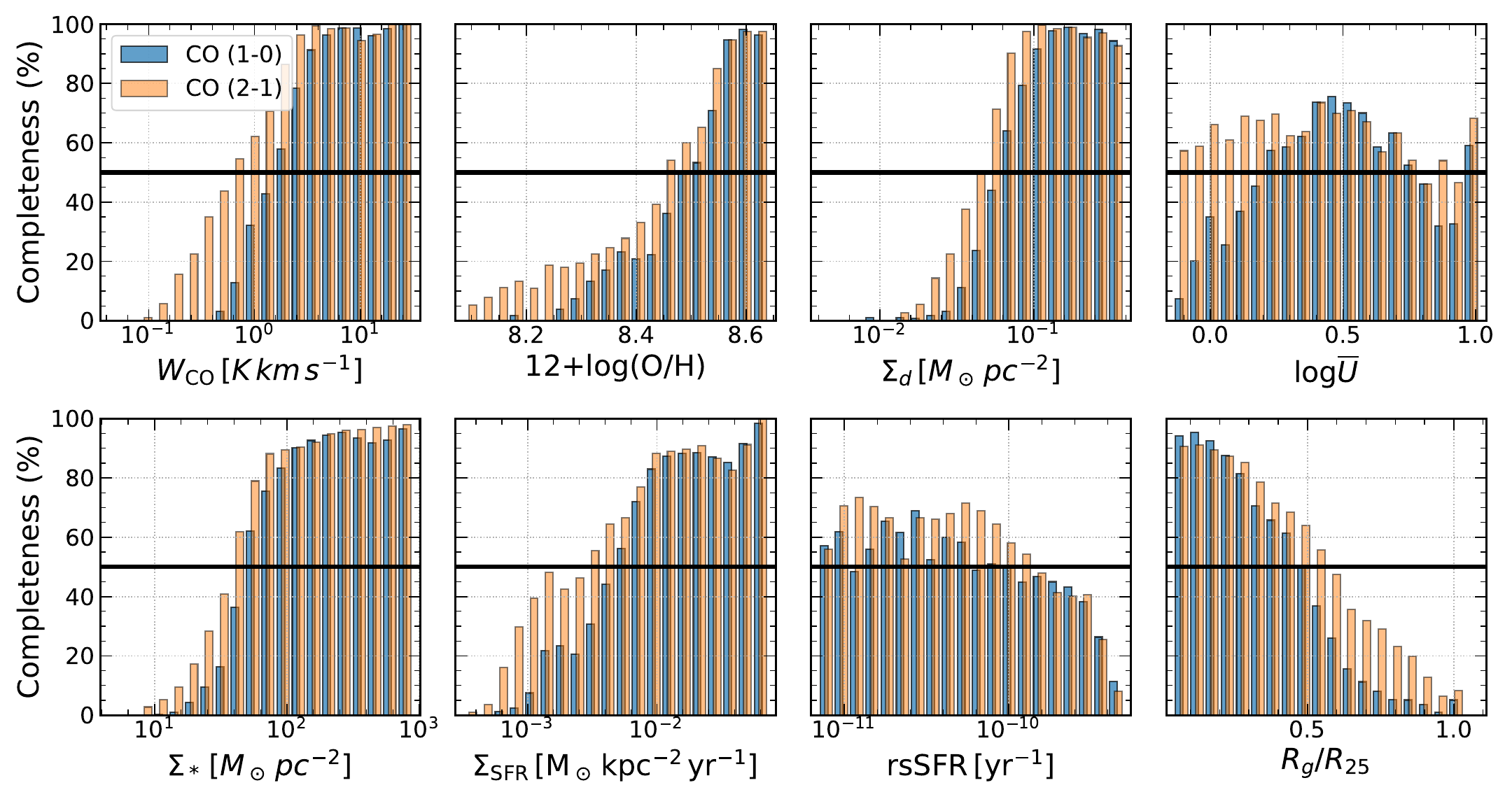}
    \caption{The completeness of our data set at each \aco-quantity pair. The $50\%$ completeness is marked with a horizontal black line. In the statistical calculations, e.g.\ linear regression and correlation, we only use  data in the parameter range with completeness\,$>50\%$.}
    \label{fig:aCO10_completeness}
\end{figure*}

\noindent\textbf{Completeness.} For statistical quantities that involve \aco and another quantity ($X$), e.g.\ correlations and linear regression, in addition to the S/N mask, we only calculate with data with high ($>50\%$) completeness in $X$ as the trend confidence criteria. The completeness for data with $X_i \leq X < X_f$, or $[X_i,~X_f)$, is defined as:
\begin{equation}
    Completeness \equiv \frac{{\rm num~of~S/N~>~1}~pixels~with~[X_i,~X_f)}{{\rm num~of~all~pixels~with}~[X_i,~X_f)},
\end{equation}
where the definition of ``S/N~>~1'' is the same as the S/N mask earlier this subsection. We show the completeness and the $50\%$ threshold in our data set in \autoref{fig:aCO10_completeness}. For most quantities, the \cotwo data has a better completeness coverage than the \coone data. Note that at the high-\Ubar end, the completeness fluctuates around 50\%. We treat all data with $\log\Ubar>0.75$ as incomplete for simplicity.

\section{Results}\label{sec:results}
% Subsection (0): Descirbe the general stats of measured \aco
\begin{table*}
\caption{Statistics of \aco measurements. All the \aco values are in unit of [\acoUnit].}\label{tab:aco_stats}
\begin{center}
\begin{tabular}{lcccccc|cccccc}
\toprule
\multirow{2}{*}{Galaxy} & \multicolumn{6}{c|}{\acotwo} & \multicolumn{6}{c}{\acoone} \\
 & Mean & W-mean$^{(a)}$ & 16$^\mathrm{th}$--84$^\mathrm{th}$ \%tile$^{(b)}$ & $W_\mathrm{CO}$\%$^{(c)}$ & N$_\mathrm{pix}^{(d)}$ & N$_\mathrm{pix,100}^{(e)}$ & Mean & W-mean$^{(a)}$ & 16$^\mathrm{th}$--84$^\mathrm{th}$ \%tile$^{(b)}$ & $W_\mathrm{CO}$\%$^{(c)}$ & N$_\mathrm{pix}^{(d)}$ & N$_\mathrm{pix,100}^{(e)}$ \\
\midrule
IC0342 & \nodata & \nodata & \nodata & \nodata & \nodata & \nodata & 2.62 & 2.31 & 1.46--4.41 & 100.0\% & 168 & 28 \\
NGC0253 & 14.15 & 5.16 & 3.55--28.62 & 99.7\% & 108 & 37 & 2.47 & 1.99 & 1.26--4.25 & 100.0\% & 39 & 31 \\
NGC0300 & 13.42 & 13.41 & 5.66--33.33 & 100.0\% & 5 & 0 & 6.64\dag & 6.63\dag & 2.74--15.59\dag & 100.0\% & 5\dag & 0\dag \\
NGC0598 & 10.58 & 10.05 & 4.29--24.69 & 65.2\% & 41 & 0 & 5.93\dag & 5.76\dag & 2.44--13.76\dag & 65.2\% & 41\dag & 0\dag \\
NGC0628 & 6.69 & 6.63 & 4.25--10.17 & 100.0\% & 172 & 57 & 3.31 & 3.32 & 1.68--6.24 & 66.8\% & 172 & 57 \\
NGC2841 & \nodata & \nodata & \nodata & \nodata & \nodata & \nodata & 9.51 & 9.15 & 3.73--22.03 & 75.5\% & 131 & 92 \\
NGC3184 & 4.12 & 4.11 & 2.09--7.68 & 57.2\% & 267 & 56 & 2.44 & 2.41 & 1.13--4.95 & 87.3\% & 247 & 56 \\
NGC3198 & 4.92 & 4.37 & 1.7--12.59 & 22.4\% & 19 & 0 & 2.63 & 2.28 & 0.81--7.49 & 26.6\% & 18 & 0 \\
NGC3351 & 7.29 & 5.41 & 3.94--11.71 & 100.0\% & 82 & 54 & 2.92 & 2.84 & 1.49--5.63 & 62.7\% & 81 & 54 \\
NGC3521 & 8.68 & 6.7 & 3.58--17.2 & 92.8\% & 206 & 143 & 4.06 & 3.67 & 1.87--8.0 & 96.0\% & 162 & 125 \\
NGC3596 & 11.14 & 9.99 & 6.19--18.28 & 100.0\% & 40 & 16 & 7.81\dag & 7.32\dag & 4.59--12.63\dag & 100.0\% & 40\dag & 16\dag \\
NGC3621 & 5.53 & 5.45 & 2.99--10.05 & 98.6\% & 52 & 19 & 3.96\dag & 4.01\dag & 2.15--7.19\dag & 98.6\% & 52\dag & 19\dag \\
NGC3627 & 4.63 & 3.84 & 2.58--7.37 & 100.0\% & 176 & 143 & 1.87 & 1.71 & 1.08--2.96 & 75.9\% & 176 & 143 \\
NGC3631 & 8.82 & 5.1 & 2.53--21.84 & 29.2\% & 102 & 17 & 3.1 & 2.55 & 1.08--7.81 & 25.3\% & 97 & 17 \\
NGC3938 & 5.91 & 5.61 & 2.78--11.7 & 71.0\% & 229 & 88 & 3.51 & 3.39 & 1.58--7.33 & 78.3\% & 224 & 88 \\
NGC3953 & 15.35 & 12.27 & 5.93--32.15 & 91.7\% & 483 & 98 & 8.27\dag & 7.19\dag & 3.42--17.22\dag & 91.7\% & 483\dag & 98\dag \\
NGC4030 & \nodata & \nodata & \nodata & \nodata & \nodata & \nodata & 5.71 & 4.57 & 2.68--11.09 & 93.8\% & 207 & 135 \\
NGC4051 & 13.85 & 9.79 & 6.08--24.27 & 87.9\% & 394 & 42 & 5.44 & 4.74 & 2.33--11.2 & 72.2\% & 200 & 36 \\
NGC4207 & 4.25 & 4.22 & 2.2--8.1 & 100.0\% & 4 & 4 & 3.32\dag & 3.31\dag & 1.77--6.16\dag & 100.0\% & 4\dag & 4\dag \\
NGC4254 & 3.93 & 4.0 & 2.37--6.43 & 90.3\% & 193 & 89 & 2.45 & 2.52 & 1.34--4.19 & 77.7\% & 193 & 89 \\
NGC4258 & \nodata & \nodata & \nodata & \nodata & \nodata & \nodata & 2.18 & 1.73 & 0.77--5.24 & 69.6\% & 56 & 55 \\
NGC4321 & 7.37 & 5.39 & 4.0--11.94 & 100.0\% & 459 & 199 & 4.45 & 3.77 & 2.64--6.75 & 99.9\% & 286 & 198 \\
NGC4450 & 8.28 & 5.27 & 2.28--18.6 & 90.0\% & 144 & 118 & 3.81\dag & 2.78\dag & 1.19--8.46\dag & 90.0\% & 144\dag & 118\dag \\
NGC4501 & \nodata & \nodata & \nodata & \nodata & \nodata & \nodata & 7.58 & 6.1 & 3.44--14.25 & 98.8\% & 336 & 235 \\
NGC4536 & 4.6 & 2.0 & 1.11--11.05 & 64.8\% & 35 & 21 & 2.08 & 1.77 & 0.8--4.01 & 44.7\% & 28 & 21 \\
NGC4569 & 7.49 & 3.98 & 2.41--13.96 & 100.0\% & 139 & 83 & 4.14 & 2.88 & 1.47--8.3 & 82.7\% & 127 & 83 \\
NGC4625 & 5.79 & 5.64 & 2.16--15.71 & 18.5\% & 5 & 0 & 3.59\dag & 3.52\dag & 1.29--9.49\dag & 18.5\% & 5\dag & 0\dag \\
NGC4651 & 4.22 & 4.15 & 2.16--7.59 & 54.4\% & 39 & 39 & 3.36\dag & 3.34\dag & 1.7--6.04\dag & 54.4\% & 39\dag & 39\dag \\
NGC4689 & 10.23 & 8.48 & 4.75--19.72 & 99.7\% & 132 & 40 & 4.59 & 4.24 & 2.09--9.21 & 73.3\% & 124 & 40 \\
NGC4725 & 12.68 & 10.56 & 4.16--29.68 & 75.7\% & 317 & 146 & 6.21\dag & 5.24\dag & 2.01--14.55\dag & 75.7\% & 317\dag & 146\dag \\
NGC4736 & 2.31 & 1.87 & 0.83--5.25 & 83.7\% & 30 & 30 & 1.26 & 1.18 & 0.49--3.01 & 100.0\% & 14 & 14 \\
NGC4941 & 8.1 & 7.1 & 3.1--16.19 & 69.5\% & 47 & 31 & 4.95\dag & 4.75\dag & 2.15--9.68\dag & 69.5\% & 47\dag & 31\dag \\
NGC5055 & 8.94 & 7.47 & 4.13--17.65 & 92.3\% & 312 & 120 & 4.55 & 4.43 & 2.46--8.24 & 100.0\% & 157 & 112 \\
NGC5248 & 11.23 & 7.02 & 4.88--19.99 & 100.0\% & 218 & 87 & 5.08 & 4.06 & 2.39--9.52 & 88.2\% & 196 & 87 \\
NGC5457 & 6.87 & 6.19 & 3.47--12.67 & 87.9\% & 419 & 50 & 3.38 & 3.16 & 1.81--6.06 & 95.6\% & 311 & 50 \\
NGC6946 & 3.27 & 2.74 & 1.7--5.91 & 87.4\% & 372 & 121 & 2.26 & 1.89 & 1.14--4.05 & 96.8\% & 312 & 121 \\
NGC7331 & 19.48 & 11.17 & 6.36--42.52 & 88.3\% & 345 & 106 & 6.41 & 5.35 & 2.67--13.69 & 86.2\% & 236 & 105 \\
Overall & 9.32 & 5.69 & 3.91--13.96 & \nodata & 5586 & 2054 & 4.22 & 3.33 & 2.21--6.09 & \nodata & 4298 & 2072 \\
Overall\dag & \nodata & \nodata & \nodata & \nodata & \nodata & \nodata & 4.72\dag & 3.48\dag & 2.32--7.23\dag & \nodata & 5475\dag & 2543\dag \\
\bottomrule
\end{tabular}
\end{center}
\tablecomments{(a) CO-intensity-weighted mean. (b) The percentiles are calculated with non-weighted data. (c) Fraction of $W_\mathrm{CO}$ recovered (above the S/N mask) in each galaxy. Galaxies with CO recovery fraction $<50\%$ will be visualized differently in figures showing galaxy-averaged values. (d) Number of pixel-by-pixel measurements. (e) Number of pixel-by-pixel with valid \aco measurements at $\Sigmastar \geq 100~\SigmaMassUnit$. This will be discussed later in \autoref{sec:aCO-SigmaMstar}. (\dag) \acoone calculated with $R_{21}(I_\mathrm{W4})$ (\autoref{eq: R21(W4)}) and \wcotwo due to no \wcoone data.}
\end{table*}

\begin{figure}
    \centering
	\includegraphics[width=1.0\columnwidth]{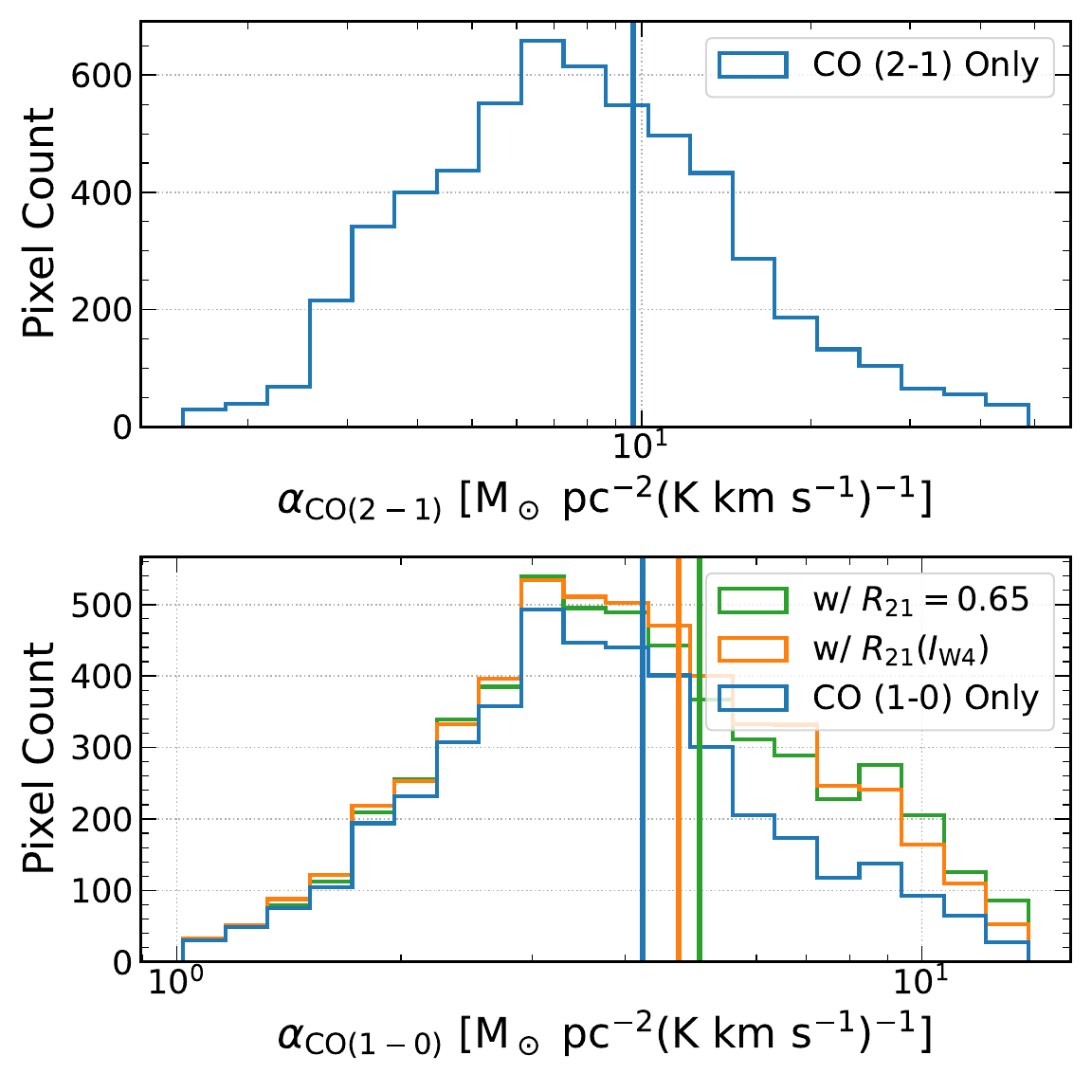}
    \caption{The distribution of measured $\acotwo$ and $\acoone$. The mean value of each type of measurement is marked in vertical lines with the corresponding color.
    }
    \label{fig:aCO_hist}
\end{figure}

% Describe stats (tab) and correlation (fig)
In total, we measure resolved \aco values across \nsample galaxies, including $\sim790$ and $\sim610$ independent measurements\footnote{In our data products, the pixel size is 1/3 of the FWHM of the Gaussian PSF. Thus, the number of independent measurements is smaller than the number of pixels listed in \autoref{tab:aco_stats}.} from \cotwo and \coone data, respectively.
We summarize the measurements in \autoref{tab:aco_stats} and the distribution of \aco in \autoref{fig:aCO_hist}. For each galaxy, we report the mean, CO-intensity-weighted mean, 16$^{\rm th}$-84$^{\rm th}$ percentiles, and number of pixel-by-pixel measurements of \aco.

The mean value for $\acotwo$ and $\acoone$ are 9.3 and 4.2~\acoUnit with 16$^{\rm th}$-84$^{\rm th}$ percentiles spanning 3.9--14.0 and 2.2--6.1~\acoUnit, respectively. The mean $\acoone$ corresponds to $\sim0.97 \acoMW$, whereas the mean $\acotwo$ corresponds to $\sim1.39 \acoMW$ assuming a constant $R_{21}=0.65$.

Besides the simple mean, we also calculate the $W_{\rm CO}$-weighted mean, which better reflects the \aco value to be adopted for data at coarser resolution. The $W_{\rm CO}$-weighted mean for $\acotwo$ and $\acoone$ are 5.69 and 3.33, respectively. Both values are lower than the simple mean, which indicates that the \aco values are lower in brighter regions. Unless specified otherwise, we use $W_\mathrm{CO}$-weighted mean \aco for galaxy-integrated analysis in the following content.

When we include \cotwo data for galaxies without \coone measurements with a variable $R_{21}$ (values with \dag\ in \autoref{tab:aco_stats}), the mean and $W_{\rm CO}$-weighted mean of $\acoone$ increase to 4.72 and 3.48, respectively, indicating that galaxies without \coone measurements in this data set tend to have larger \aco. This is also visualized in the bottom panel of \autoref{fig:aCO_hist}. Also, we find that the distribution of $\acoone$ does not differ much between the two $R_{21}$ prescriptions adopted in this work: the fixed $R_{21}=0.65$ and the $I_{\rm W4}$-dependent $R_{21}$.

In \autoref{tab:aco_stats}, we also list the $W_\mathrm{CO}$ recovery fraction ($W_\mathrm{CO}\%$), which is the percentage of $W_\mathrm{CO}$ recovered (above the S/N mask) in the pixel-by-pixel analysis in each galaxy. We notice a few galaxies with low $W_\mathrm{CO}$ recovery fraction, meaning that there are significant numbers of pixels with $W_\mathrm{CO}$ detection removed from analysis. The main reason for NGC3631 and most galaxies with recovery fraction above 50\% is low sensitivity in dust/IR data. In NGC3631, $>85\%$ of pixels removed have ${\rm S/N} < 1$ in the IR bands. In NGC3198 and NGC4625, the sensitivity in dust/IR only explains $<60\%$ of pixels removed. The rest of the pixels were removed due to ${\rm S/N} < 1$ in \Sigmamol, a combined effect of \Sigmad, \Sigmaatom and \metal. This type of pixels have ${\rm S/N} > 1$ in $W_\mathrm{CO}$ and ${\rm S/N} < 1$ in \Sigmamol, likely indicating a small \aco.

% Subsection: How \aco correlates with other parameters
\subsection{Correlations with Local Conditions}\label{sec:results:local}

\begin{table}
\caption{The correlation and linear regression between pixel-by-pixel \aco measurements and local physical quantities.}
\begin{center}
\label{tab: aCO_prop}
\begin{tabular}{lc|ccc|c}
\toprule
\multicolumn{6}{c}{$\log\acotwo$, \cotwo Only}\\
\midrule
$x^{(1)}$ & $\rho$ & $m^{(1)}$ & $b^{(1)}$ & \rms & $\rho_\mathrm{norm}^{(2)}$ \\
\midrule
$\log W_\mathrm{CO}$ & -0.67  & -0.49$\pm$0.01 & 1.01$\pm$0.0 & 0.17 & -0.33  \\
\metal & -0.48  & -2.62$\pm$0.06 & 23.2$\pm$0.5 & 0.22 & -0.2  \\
$\log\Sigmad$ & -0.44  & -0.66$\pm$0.01 & 0.15$\pm$0.01 & 0.24 & -0.29  \\
$\log\Ubar$ & -0.73  & -0.97$\pm$0.01 & 1.15$\pm$0.0 & 0.19 & -0.4  \\
$\log\Sigmastar$ & -0.32  & -0.38$\pm$0.01 & 1.57$\pm$0.02 & 0.23 & \underline{-0.4}  \\
$\log\Sigmasfr$ & -0.59  & -0.48$\pm$0.01 & -0.18$\pm$0.01 & 0.17 & -0.3  \\
$\log$\,sSFR & -0.22  & -0.12$\pm$0.01 & -0.4$\pm$0.1 & 0.27 & 0.1  \\
$R_g/R_{25}$ & 0.43  & 0.8$\pm$0.02 & 0.56$\pm$0.01 & 0.23 & 0.38  \\
\toprule
\multicolumn{6}{c}{$\log\acoone$, \coone Only}\\
\midrule
$x$ & $\rho$ & $m$ & $b$ & \rms & $\rho_\mathrm{norm}^{(2)}$ \\
\midrule
$\log W_\mathrm{CO}$ & -0.42  & -0.39$\pm$0.01 & 0.83$\pm$0.01 & 0.18 & \underline{-0.43}  \\
\metal & -0.32  & -1.81$\pm$0.08 & 16.0$\pm$0.7 & 0.2 & -0.17  \\
$\log\Sigmad$ & -0.07  & -0.14$\pm$0.02 & 0.39$\pm$0.02 & 0.21 & \underline{-0.22}  \\
$\log\Ubar$ & -0.35  & -0.71$\pm$0.02 & 0.8$\pm$0.01 & 0.19 & -0.23  \\
$\log\Sigmastar$ & -0.04  & -0.12$\pm$0.01 & 0.76$\pm$0.02 & 0.23 & \underline{-0.3}  \\
$\log\Sigmasfr$ & -0.31  & -0.29$\pm$0.01 & -0.05$\pm$0.02 & 0.19 & -0.24  \\
$\log$\,sSFR & -0.26  & -0.22$\pm$0.02 & -1.7$\pm$0.2 & 0.22 & 0.1  \\
$R_g/R_{25}$ & 0.25  & 0.56$\pm$0.03 & 0.36$\pm$0.01 & 0.21 & \underline{0.29}  \\
\toprule
\multicolumn{6}{c}{$\log\acoone$, w/ $R_{21}(I_\mathrm{W4})$}\\
\midrule
$x$ & $\rho$ & $m$ & $b$ & \rms & $\rho_\mathrm{norm}^{(2)}$\\
\midrule
$\log W_\mathrm{CO}$ & -0.46  & -0.41$\pm$0.01 & 0.84$\pm$0.01 & 0.18 & -0.29  \\
\metal & -0.4  & -2.19$\pm$0.07 & 19.3$\pm$0.6 & 0.21 & -0.19  \\
$\log\Sigmad$ & -0.17  & -0.25$\pm$0.02 & 0.31$\pm$0.02 & 0.22 & -0.17  \\
$\log\Ubar$ & -0.34  & -0.7$\pm$0.02 & 0.8$\pm$0.01 & 0.19 & -0.19  \\
$\log\Sigmastar$ & -0.15  & -0.18$\pm$0.01 & 0.92$\pm$0.02 & 0.23 & \underline{-0.3}  \\
$\log\Sigmasfr$ & -0.31  & -0.31$\pm$0.01 & -0.07$\pm$0.02 & 0.19 & -0.22  \\
$\log$\,sSFR & -0.2  & -0.08$\pm$0.01 & -0.3$\pm$0.1 & 0.24 & 0.16  \\
$R_g/R_{25}$ & 0.31  & 0.64$\pm$0.03 & 0.36$\pm$0.01 & 0.21 & 0.28  \\
\toprule
\multicolumn{6}{c}{$\log\acoone$, w/ $R_{21}=0.65$}\\
\midrule
$x$ & $\rho$ & $m$ & $b$ & \rms & $\rho_\mathrm{norm}^{(2)}$\\
\midrule
$\log W_\mathrm{CO}$ & -0.45  & -0.4$\pm$0.01 & 0.84$\pm$0.01 & 0.18 & -0.28  \\
\metal & -0.4  & -2.29$\pm$0.08 & 20.1$\pm$0.7 & 0.22 & -0.18  \\
$\log\Sigmad$ & -0.16  & -0.25$\pm$0.02 & 0.31$\pm$0.02 & 0.22 & -0.15  \\
$\log\Ubar$ & -0.37  & -0.73$\pm$0.02 & 0.82$\pm$0.01 & 0.19 & -0.2  \\
$\log\Sigmastar$ & -0.16  & -0.19$\pm$0.01 & 0.95$\pm$0.02 & 0.24 & \underline{-0.3}  \\
$\log\Sigmasfr$ & -0.32  & -0.3$\pm$0.01 & -0.07$\pm$0.02 & 0.19 & -0.22  \\
$\log$\,sSFR & -0.26  & -0.14$\pm$0.01 & -0.8$\pm$0.1 & 0.26 & 0.15  \\
$R_g/R_{25}$ & 0.32  & 0.68$\pm$0.03 & 0.35$\pm$0.01 & 0.22 & 0.28  \\
\bottomrule
\end{tabular}
\end{center}
\tablecomments{All correlation coefficients presented have their $p$-value smaller than 0.05. (1) The linear regression formula is $\log\aco=mx+b$. An uncertainty of $\pm 0.0$ represents that the rounded uncertainty of the parameter is smaller than $0.01$. (2) Correlation coefficients calculated with \aco normalized by $W_\mathrm{CO}$-weighted mean value in each galaxy. We underline the cases where the correlation of normalized \aco is stronger than the one without normalization.}
\end{table}

\begin{figure*}
    \centering
	\includegraphics[width=0.99\textwidth]{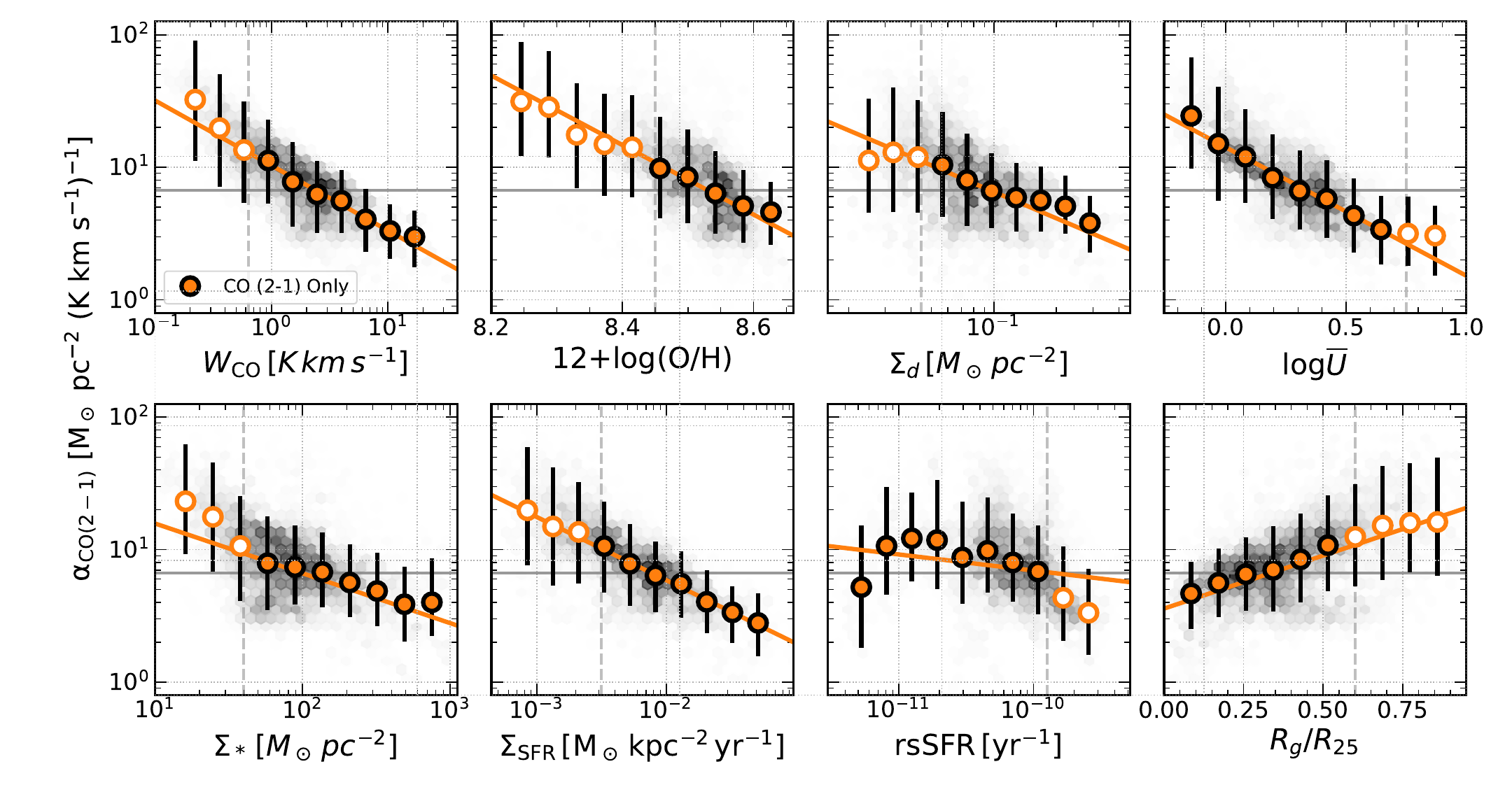}
    \caption{Our measured $\acotwo$, measured with \wcotwo data only, as a function of environmental parameters.
    The orange circles show the median of binned data. The errorbars include both the scatter within a bin and uncertainties of pixel-by-pixel measurements. The orange line shows the linear regression between $\acotwo$ and the quantity on the horizontal axis.
    The empty circles indicate bins where the quantity in the horizontal axis is incomplete, i.e.\ less than 50\% of the pixels have S/N > 1 in $W_\mathrm{CO}$ and \textit{Herschel} bands (see \autoref{sec:data:masking and completeness}). The dashed vertical line shows the 50\% boundary.
    The gray shaded region shows the hexagonal-binned pixel-by-pixel measurements.
    The gray horizontal line shows the value of \acoMW, converted to $\acotwo$ assuming $R_{21}=0.65$.
    }
    \label{fig:aCO21_properties}
\end{figure*}

\begin{figure*}
    \centering
	\includegraphics[width=0.99\textwidth]{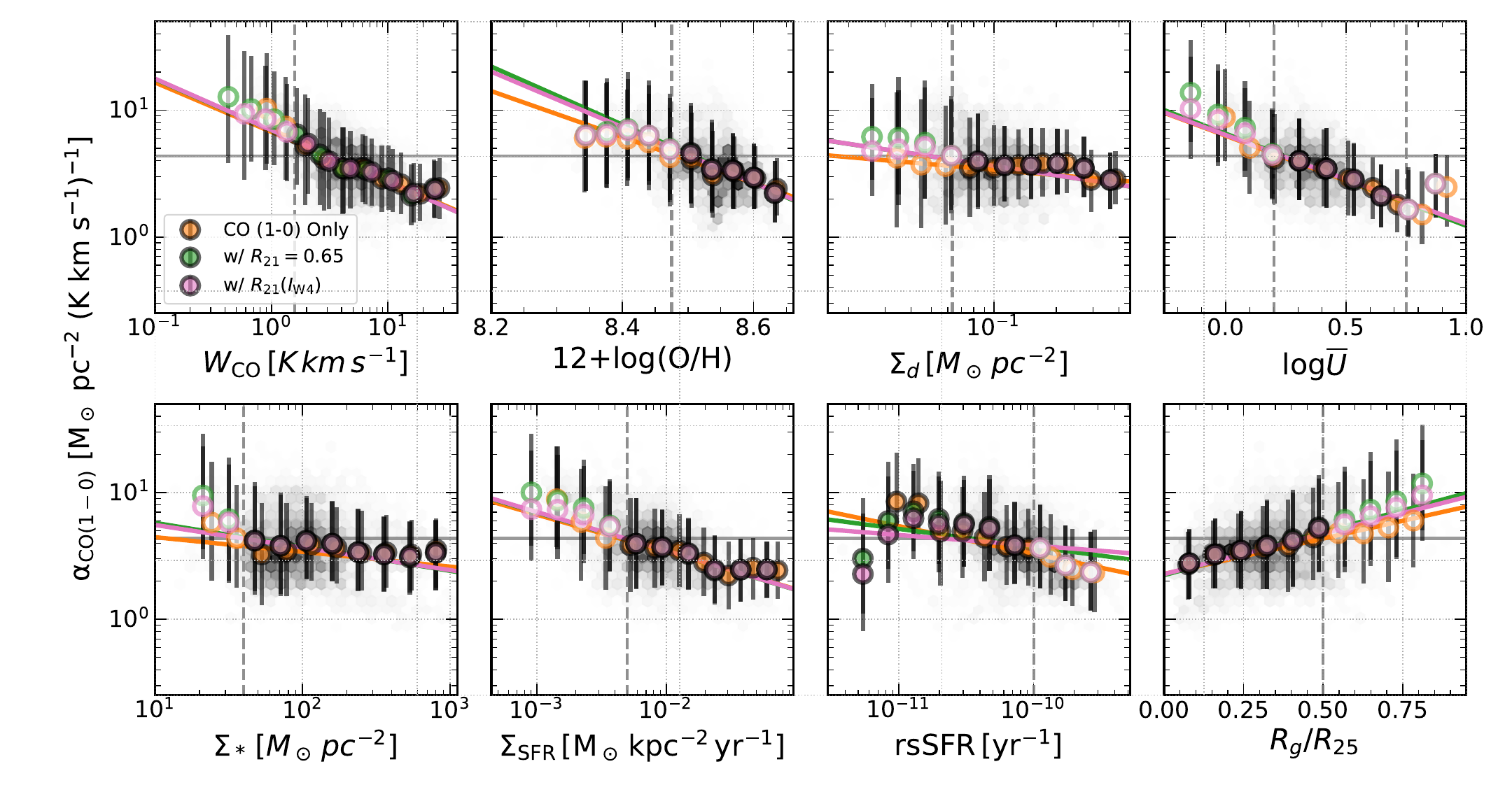}
    \caption{Similar with \autoref{fig:aCO21_properties}, but for $\acoone$ measurements. The three colors of the circles and lines show the three cases indicated in the legend. \textit{\coone Only}: measured with \wcoone data only; \textit{w/ $R_{21}=0.65$} and \textit{w/ $R_{21}(I_\mathrm{W4})$}: measured with \wcoone and \wcotwo data for galaxies without \wcoone. The background graded region is plotted for the ``\coone Only'' measurements.
    }
    \label{fig:aCO10_properties}
\end{figure*}

% Opening
We measure the pixel-by-pixel correlations between \aco and several parameters describing local physical conditions. These results are summarized in \autoref{tab: aCO_prop} and visualized in \autoref{fig:aCO21_properties}-\autoref{fig:aCO10_properties}. Note that the correlations and linear regressions\footnote{The regression for most quantities are done in logarithmic scale. See \autoref{tab: aCO_prop}.} are calculated with data in the complete zone only (with data completeness~$>50\%$; see \autoref{sec:data:masking and completeness} for details). The errorbars in \autoref{fig:aCO21_properties}-\autoref{fig:aCO10_properties} include both the scatter within a bin and the uncertainties of pixel-by-pixel measurements. We first bootstrap the measurements by 1000 times with uncertainties, and then sample the 16th- and 84-percentiles in each bin from the bootstrapped sample as the errorbars. We apply the same method for visualizing the other binned data in this work.

As shown in \autoref{tab: aCO_prop}, most quantities have significant correlations ($p$-value < 0.05) with \aco except \Sigmastar. \Sigmastar has significant correlations with $\acotwo$ from \cotwo data only and $\acoone$ from \coone only, but not when we combine \cotwo and \coone data. This is likely due to the negative \Sigmastar-to-$\acotwo$ and the positive \Sigmastar-to-$\acoone$ correlations, although both of which are weak.

$\log\Ubar$ has the strongest negative correlation with $\acotwo$, meaning that $\acotwo$ decreases toward regions with stronger interstellar radiation field strength. This is consistent with the picture that \aco decreases with higher gas temperature and larger line width \citep{BOLATTO13}. It is also the case that a higher $\log\Ubar$ might correspond to a lower \Sigmad as a caveat of our fitting methodology \citep[equivalent to ``fixing $\beta$'' in modified blackbody models, see][]{SHETTY09,SHETTY09B}. However, since we do not see a strong \Sigmad-to-$\acotwo$ correlation, the above effect should be minor. Several other quantities also show moderate (negative) correlations with $\acotwo$, i.e.\ \Sigmasfr and rsSFR. Studies have shown strong correlations between $\log\Ubar$ and \Sigmasfr \citep[e.g.][J. Chastenet et al. in preparation]{Hirashita_Chiang22,Chiang23}. Another quantity that shows moderate correlation is $W_\mathrm{CO}$. \aco is expected to anti-correlate with $W_\mathrm{CO}$ due to either external pressure or other dynamical effects \citep[e.g.][]{BOLATTO13,Sun18}. The power-law index for $W_\mathrm{CO}$ is within the previously reported range of $-0.32$ to $-0.54$ \citep{NARAYANAN12,Gong20,Hunt23}.

% (continue) strong correlations for alpha\coone
Compared to $\acotwo$, the correlations between $\acoone$ and local conditions are overall weaker. $\log~{\rm rsSFR}$ has the strongest correlation with $\acoone$ in all three $R_{21}$ cases, and $\log\Ubar$ is the second strongest.

% (continue) linear regression results
For all combinations of $\aco$ and local conditions, we perform linear regression with the functional form:
\begin{equation}
    \log\aco=mx+b,
\end{equation}
where $x$\footnote{Note that most $x$ quantities are in logarithmic scale.} is the local condition and both $m$ and $b$ are free parameters, all of which are listed in \autoref{tab: aCO_prop}. In the same table, we also report the root-mean-square error (\rms) between the measured and fitted $\log\aco$. Most formulas have $\rms\sim 0.2$~dex. $W_\mathrm{CO}$, \Ubar and \Sigmasfr have the strongest correlations and smallest \rms in general.

% Other quantities we want to talk about? e.g. metallicity, stellar mass? (can change order)
One quantity that is often used for parametrizing \aco is \metal. In our measurements, \aco has moderate to weak correlation with \metal in all cases. The slope from linear regression ($m$) is $-2.6$ for $\acotwo$ and $-1.8$ to $-2.3$ for $\acoone$. These values are mildly steeper than most literature values \citep[$\sim -1.6$ to $-2.0$,][]{BOLATTO13,HUNT15,ACCURSO17,Sun20}, but are still within previous reported range, e.g.\ $-2.0$ to $-2.8$ in \citet{SCHRUBA12}. Note that our data set is less suitable for an in-depth study on \metal since $>80\%$ of our data is concentrated in a small dynamic range of $8.4 \leq \metal \leq 8.6$. 

% Normalize stuff
We also calculate the correlation between physical quantities and normalized \aco. In this calculation, \aco is normalized by the $W_\mathrm{CO}$-weighted mean in each galaxy. For most quantities, the correlation become weaker after normalization. Meanwhile, \Sigmastar has a stronger correlation with normalized \aco in all cases in \autoref{tab: aCO_prop}, indicating that \Sigmastar traces the intra-galaxy \aco variations after normalization of galaxy-to-galaxy differences.

% Short conclusion
Overall, we have shown that $\acotwo$ has stronger correlations with local conditions than $\acoone$. Among the local quantities, $W_\mathrm{CO}$, \Ubar, and \Sigmasfr usually have stronger correlations with \aco. We do not see a strong correlation coefficient between \aco and \metal, one of the most frequently used quantity to model \aco.

\subsection{Correlation with Galaxy-Averaged Quantities}\label{sec:results:global}

\begin{figure*}
    \centering
	\includegraphics[width=0.99\textwidth]{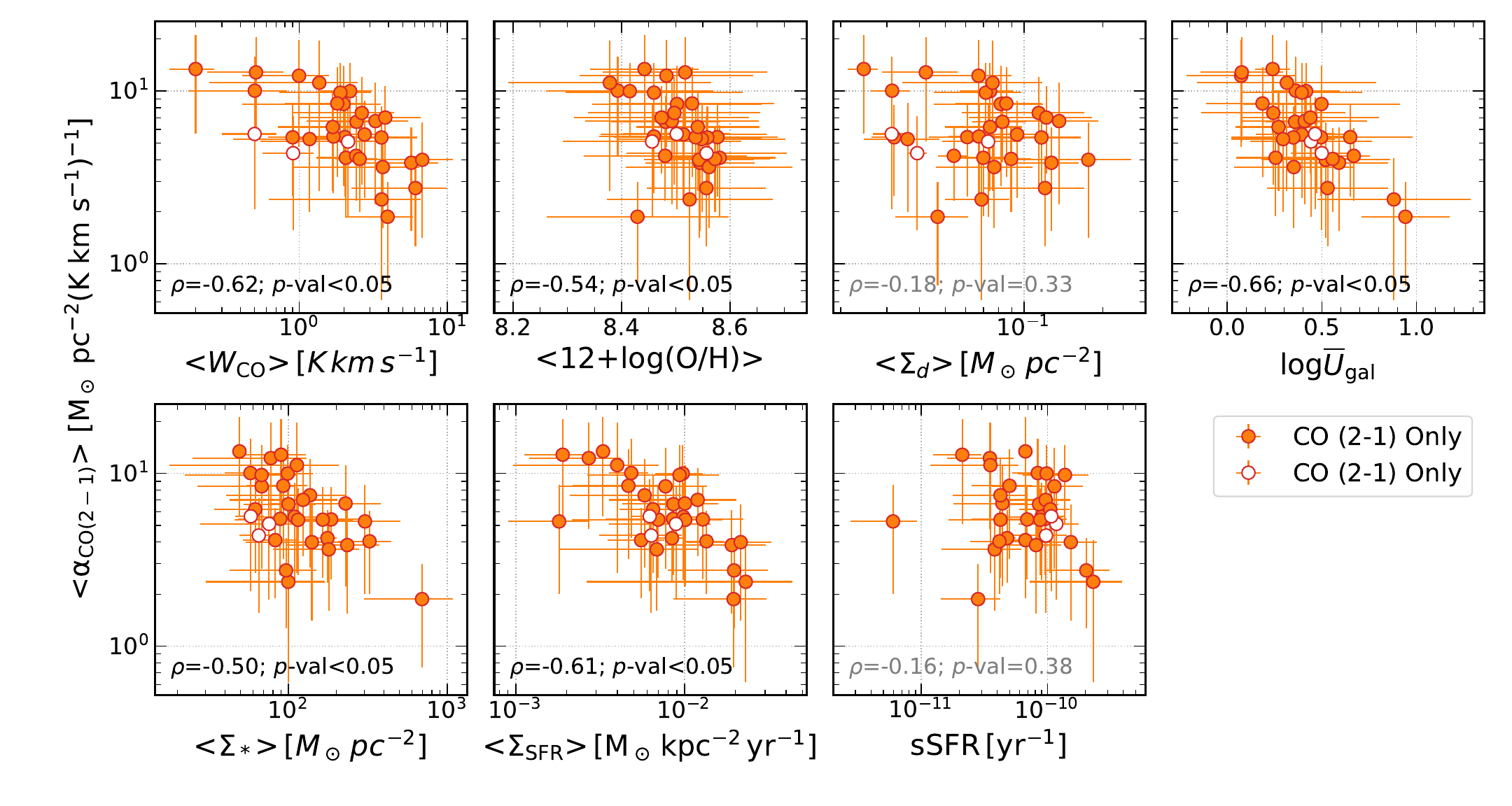}
    \caption{
    Scaling relations between galaxy-averaged $\acotwo$ and environmental parameters. The errorbars show the 16th-84th percentiles in each galaxy.
    All data points are calculated with \wcotwo data only.
    The filled markers show the galaxies with $W_\mathrm{CO}$ recovery fraction $\geq 50\%$, while the empty markers show the $<50\%$ ones (\autoref{tab:aco_stats}).
    The correlation coefficients and $p$-values are labeled at the lower left in each panel, highlighting the significant ($p$-value$<0.05$) ones. We use weighted averaged for \aco, \Ubar and sSFR, and simple averages for the other quantities. Please see \autoref{sec:results:global} for details.}
    \label{fig:mean_aCO21_properties}
\end{figure*}

\begin{figure*}
    \centering
	\includegraphics[width=0.99\textwidth]{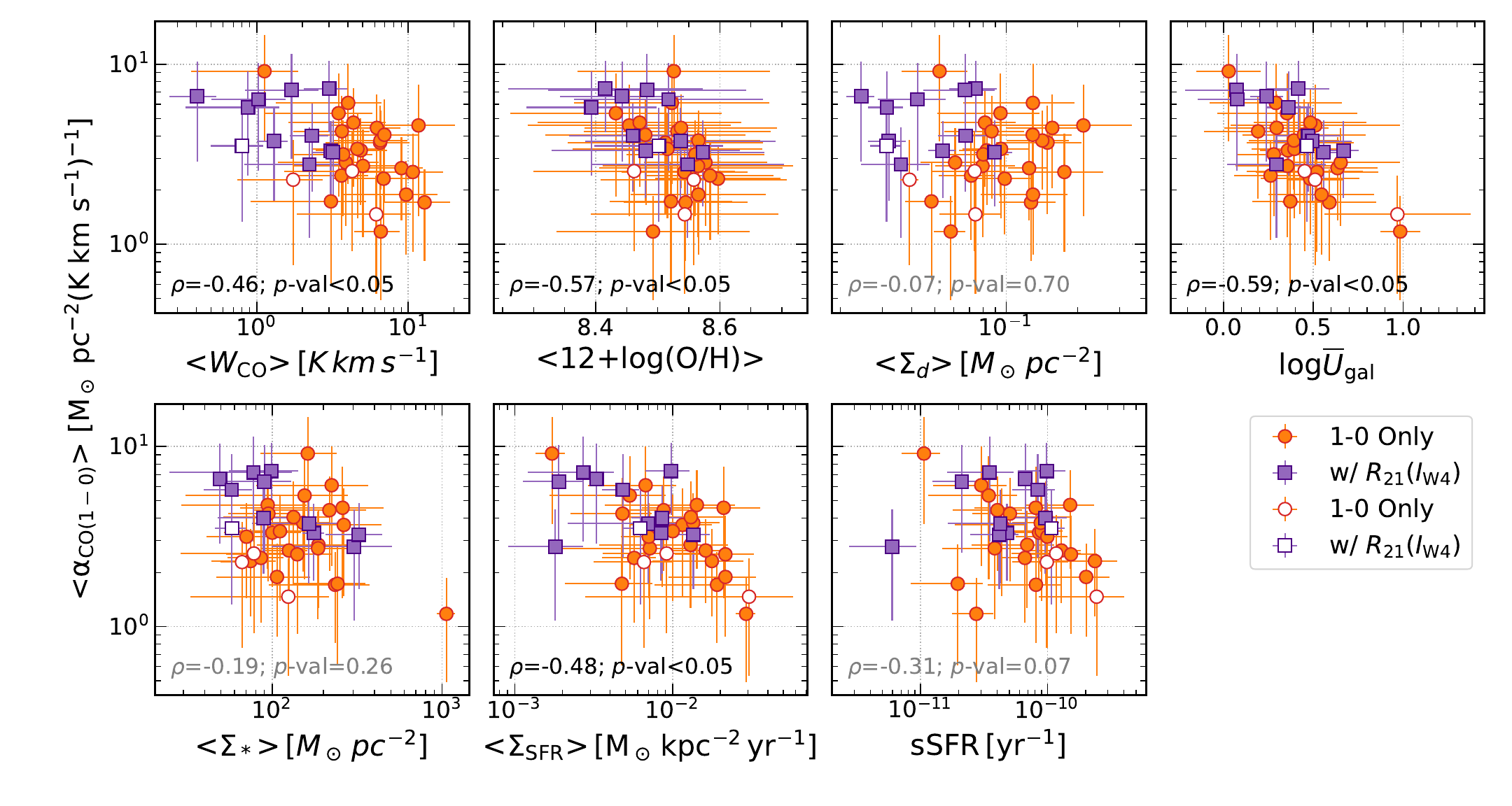}
    \caption{Similar with \autoref{fig:mean_aCO21_properties} but for $\acoone$ measurements.
    The orange points show the measurements with \wcoone data only, and the purple points show measurements with \wcotwo data with a $I_\mathrm{W4}$-dependent $R_{21}$. The fixed $R_{21}$ results are similar to the ones from $I_\mathrm{W4}$-dependent $R_{21}$. They are not displayed for clarity.
    }
    \label{fig:mean_aCO10_properties}
\end{figure*}

Besides kpc-scale variations, we also test how galaxy-averaged $\acotwo$ and $\acoone$ vary between galaxies, and how their variations correlate with galaxy-averaged properties. The results are visualized in \autoref{fig:mean_aCO21_properties}-\autoref{fig:mean_aCO10_properties}. We use the symbol $\text{<}X\text{>}$ to represent the simple averaged value of quantity $X$ in each galaxy, i.e.\ $\Sigma_i^{\rm gal} X_i / \Sigma_i^{\rm gal}$, where $\Sigma_i^{\rm gal}$ is the summation over all pixels in a galaxy. There are three quantities for which we do not apply simple averages: (1) as mentioned in \autoref{sec:results}, we will present the $W_\mathrm{CO}$-weighted mean of \aco; (2) $\log\Ubar_\mathrm{gal}$ is calculated as $\log \big(\Sigma_i^{\rm gal}\Sigma_{{\rm dust},i} \Ubar_i/\Sigma_i^{\rm gal}\Sigma_{{\rm dust},i}$\big) to reflect the dust-mass-weighted averaged ISRF; (3) sSFR is calculated as $\Sigma_i^{\rm gal}\Sigma_{{\rm SFR},i}/\Sigma_i^{\rm gal}\Sigma_{\star,i}=SFR/M_\star$.
The errorbars in \autoref{fig:mean_aCO21_properties}-\autoref{fig:mean_aCO10_properties} show the 16th-84th percentiles of the corresponding quantity. In \autoref{fig:mean_aCO10_properties}, we also include $\acoone$ calculated with \cotwo data with a $I_\mathrm{W4}$-dependent $R_{21}$. We only include one $R_{21}$ prescription here for clarity.

We report the correlation coefficients and the corresponding $p$-values in each panel of \autoref{fig:mean_aCO21_properties}-\autoref{fig:mean_aCO10_properties}. Compared to the results in \autoref{sec:results:local}, we note that whether a quantity has significant correlation with \aco and the strength of the correlation often differ between the spatially resolved case and the galaxy-averaged case. Several quantities show significant correlations with <$\alpha_\mathrm{CO}$>. \metal and <$\Sigma_\mathrm{SFR}$> seem to show stronger correlation with <$\alpha_\mathrm{CO}$> for both \cotwo and \coone than in the spatially resolved case. The insignificant correlation between <$\Sigma_{\star}$> and \acoone is consistent with the findings in \citet{Carleton17} and \citet{Dunne22} assuming that \Sigmastar dominates the total mass surface density and that \coone dominates the CO measurements.

\section{Power-law dependence of the conversion factor on
stellar mass surface density}\label{sec:aCO-SigmaMstar}

\begin{figure*}
    \centering
	\includegraphics[width=\textwidth]{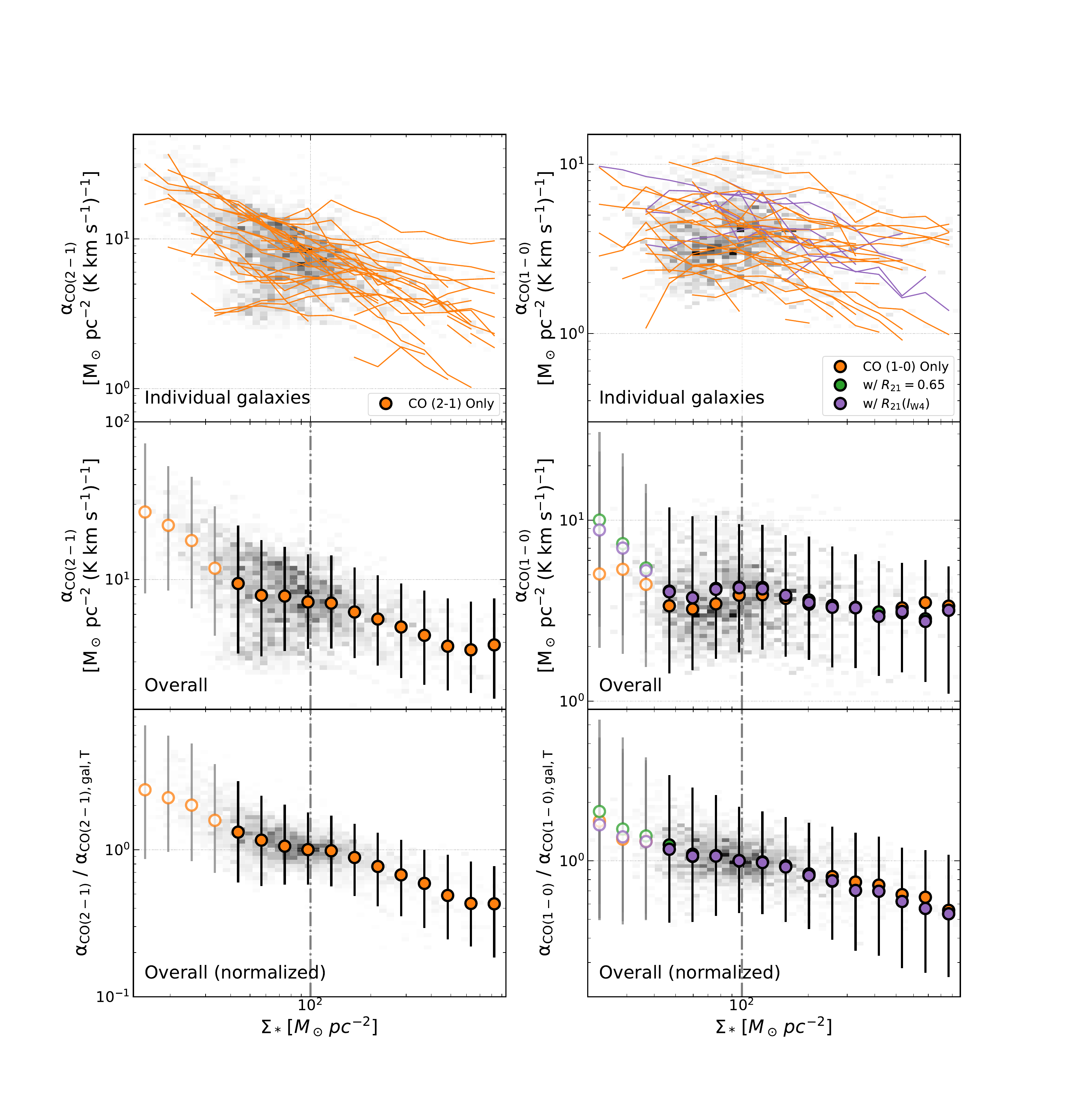}
    \caption{Measured \aco plotted against stellar mass surface density. The left panels show the $\acotwo$ measurements while the right panels show the $\acoone$ measurements.
    Top panels: the measured \aco in each galaxy, where the solid lines show the median in each \Sigmastar bin.
    Middle panels: the overall binned median of \aco, where the errorbars show the 16th- and 84th-percentiles.
    Bottom panels: similar to the middle panels, but the y-axis is normalized to the \aco value at $\SigmastarT=100~\SigmaMassUnit$ (shown in vertical dashed line) in each galaxy. The galaxies with \aco increasing with \Sigmastar are removed in the bottom panels.
    The 2d histogram in all panels shows the overall data distribution.
    }
    \label{fig:aCO_first}
\end{figure*}

In the \aco prescription proposed in \citet{BOLATTO13}, the authors use a power law with \Sigmatot (=\Sigmastar+\Sigmagas) to trace the changes in \aco due to CO emissivity variations (related to gas temperature and opacity) and a threshold in \Sigmatot to trace where the effects become important. Inspired by their work and motivated by the necessity of improving \aco prescriptions in galaxy centers, we examine whether a similar functional form applies to our \aco measurements. 
Furthermore, as shown in the previous section, the correlation between \Sigmastar and \aco improves after normalizing \aco the their $W_\mathrm{CO}$-weighted mean, which could fit into the picture of separating CO-dark and starburst components in \citet{BOLATTO13}.
With the WISE full-sky observations, the resolved \Sigmastar for all nearby galaxies is widely available, which makes this kind of prescription easy to apply. 
In this study, we will focus on the \aco-to-\Sigmastar relation instead of \Sigmatot because our data set is mostly \Sigmastar-dominated (50\% with $\Sigmagas/\Sigmatot<0.2$; 99.5\% with $\Sigmagas/\Sigmatot<0.5$).
% Not sure if we should mention "to avoid circular issue in calculating alphaCO" as a reason

We present the correlations between \aco and \Sigmastar in \autoref{fig:aCO_first} for both $\acotwo$ (left panels) and $\acoone$ (right panels).
In the top panels, we present the profile of measured \aco versus \Sigmastar in each galaxy. For $\acotwo$, most galaxies have their \aco anti-correlate with \Sigmastar at $\Sigmastar>100~\SigmaMassUnit$ aside from a few exceptions. It is similar for $\acoone$, but with a flatter \aco-to-\Sigmastar slope. At the low-\Sigmastar end, some galaxies still have negative \aco-to-\Sigmastar correlations while the others have strong positive correlations.
In the middle panels, we show the collective behavior across galaxies using a binned average as a function of \Sigmastar, with each bin spanning $\sim 0.1$~dex in \Sigmastar.

We find that in regions with high-\Sigmastar, \aco generally decreases with \Sigmastar, which is consistent with the negative power-law index in the \citet{BOLATTO13} formula. There appears to be galaxy-to-galaxy variation in the value of \aco, but good agreement in the rate of how fast $\aco$ decreases with $\Sigmastar$. 
To better illustrate this phenomena, we normalize \aco in each galaxy at a threshold $\SigmastarT\equiv 100~\SigmaMassUnit$ \citep[a threshold inspired by][]{BOLATTO13} and show the normalized \aco in the bottom panel of \autoref{fig:aCO_first}. The normalization in each galaxy (\acoT) is defined as the median \aco of pixels with their \Sigmastar within $\SigmastarT\pm 0.05$~dex.
% (the variation of \SigmastarT will be explored later in this section).

In the remainder of this section, we will focus on analyzing the scaling relation between \aco and \Sigmastar in a sub-sample of galaxies with at least 5 measurements with $\Sigmastar > \SigmastarT$ (29 galaxies for \cotwo and 25 galaxies for \coone, see the $\rm N_{pix,100}$ column in \autoref{tab:aco_stats}). We use a power law to characterize this scaling relation:
\begin{equation}\label{eq:power law}
    \log \frac{\aco}{\acoT} = a\times\log\frac{\Sigmastar}{\SigmastarT} + b,~\Sigmastar\geq\SigmastarT,
\end{equation}
where $a$ (the power-law index) and $b$ (the offset) are free parameters. Since both \aco and \Sigmastar are normalized in the formula, we expect $b\sim 0$ (and $b_\mathrm{gal}\sim 0$) if \aco monotonically decreases with \Sigmastar. By default, we fit \autoref{eq:power law} with all data. When fitting data in individual galaxies only, we will describe the parameters as $a_\mathrm{gal}$ and $b_\mathrm{gal}$.

We exclude data from galaxies that do not satisfy the following criteria:
(1) at least 5 measurements at $\Sigmastar>\SigmastarT$;
(2) spanning at least 0.1~dex in \Sigmastar at $\Sigmastar>\SigmastarT$. Since all criteria are \SigmastarT-dependent, we expect the size of sub-sample space to vary with \SigmastarT. We will visualize the galaxies not satisfying the last two criteria in \autoref{sec:aCO-SigmaMstar:galaxy}.

With the fiducial setting, i.e.\ $\mathrm{D/M}=0.55$ and $\SigmastarT=100~\SigmaMassUnit$, the fitting yields $a=-0.50^{+0.07}_{-0.06}$ and $b=0.03^{+0.01}_{-0.01}$ with $\rms=0.15~\mathrm{dex}$ for \cotwo; $a=-0.22^{+0.06}_{-0.06}$ and $b=0.003^{+0.01}_{-0.01}$ with $\rms=0.13~\mathrm{dex}$ for \coone\footnote{A recent review (E. Schinnerer \& A. K. Leroy ARA\&A submitted) indicates a slightly different result of $a\sim-0.25$ for \acoone, which is consistent with our result. The two main differences between this work and theirs are: (1) The Schinnerer \& Leroy review adopts a different formula for \Sigmasfr-dependent $R_{21}$; (2) The Schinnerer \& Leroy review combines all available \cotwo and \coone data, while we keep them separate in this section.}.
The small $b$ values, which are consistent with our expectations, indicate that the \aco-to-\Sigmastar relation matches with the picture of a simple power law. 
The difference between the $a$ values for \cotwo and \coone data is consistent with the finding that $R_{21} \propto I_\mathrm{MIR}^{-0.2}$ \citep{Leroy23_JWST-CO}.
The uncertainties of $a$ and $b$ are estimated from 1000 rounds of bootstrap resampling. In each round, we select 29 (25) galaxies for \cotwo (\coone) data with replacement from our sample galaxies to fit the power law. We then take the difference between the best-fit parameter and the 84th- (16th-) percentile from the 1000 bootstraps as the upper (lower) uncertainty.

We will measure how the \aco-to-\Sigmastar relation varies according to the adopted D/M and \SigmastarT in the remainder of this section. In \autoref{app:d/m}, we also test how the results would change with an internal variations of D/M. Our toy model in \autoref{app:d/m} shows that $a$ could be up to $\sim 0.2$ steeper than the constant D/M case.

\subsection{Dependence of the power-law index on adopted D/M and \texorpdfstring{\SigmastarT}{}}\label{sec:aCO-SigmaMstar:D/M}
% Result 3: Evolution of the power-law index \& normalization with assumed D/M. Figure~\autoref{fig:fit_evolution}.

\begin{figure}
    \centering
	\includegraphics[width=0.99\columnwidth]{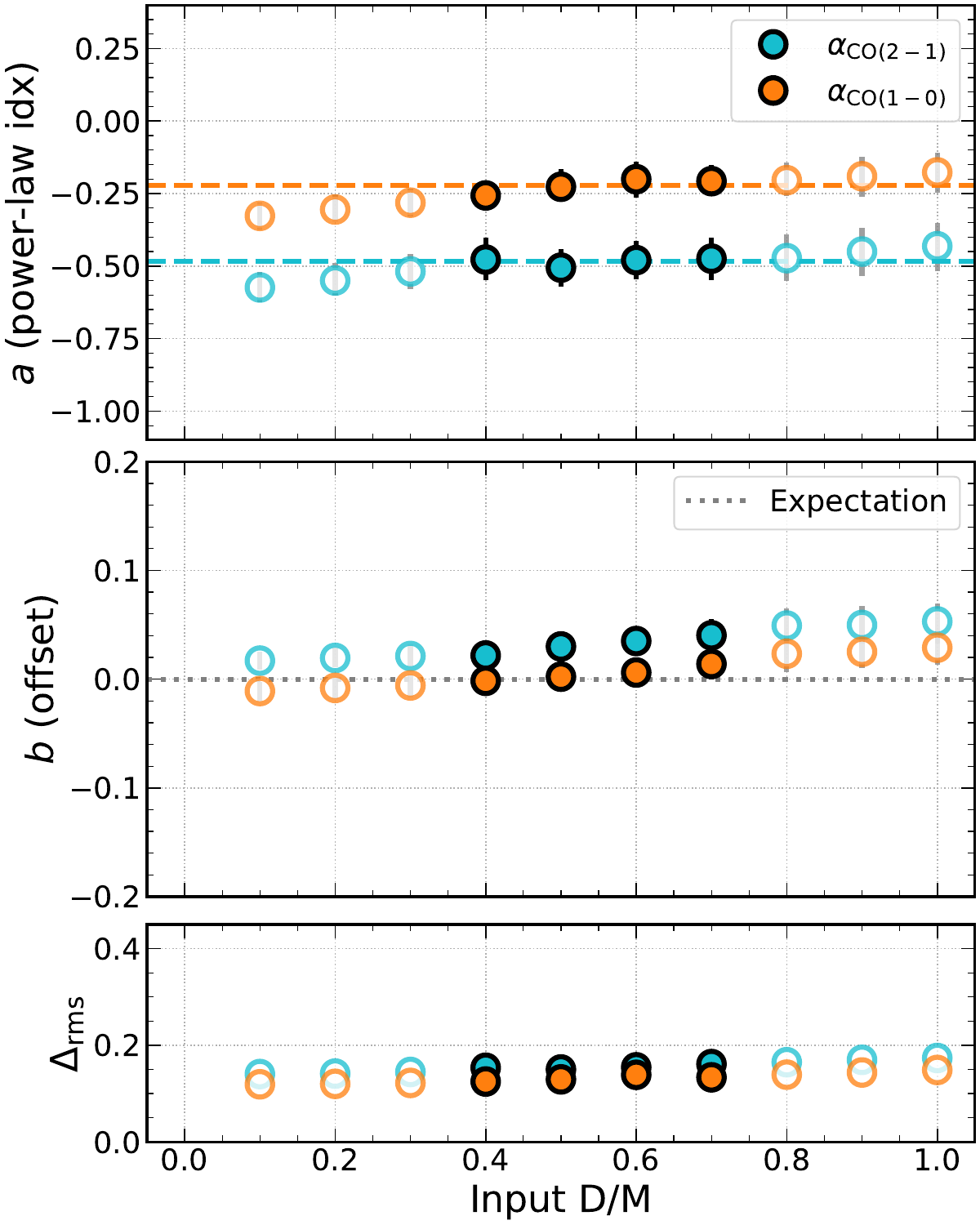}
    \caption{The dependence of fitted power-law index and offset on assumed D/M.
    The cyan points show results for \acotwo, and the orange points show results for \acoone. For \acoone, we only show results from \wcoone data only since fitting results including \wcotwo data have mininal difference.
    We highlight the region with D/M inferred from literature, i.e. $0.4\leq \mathrm{D/M} \leq 0.7$, while the empty circles show results outside that range.
    Top: the power-law index ($a$). The dashed lines show the mean value of $a$ in the range of $0.4\leq \mathrm{D/M} \leq 0.7$.
    Middle: the offset ($b$). The dotted line shows $b=0$, where expect the fitting result to be if \aco monotonically decreases with \Sigmastar.
    Bottom: the \rms of each fit.}
    \label{fig:fit_evolution_DM}
\end{figure}

To test the robustness of our results for potentially different dust properties, we expand the assumed D/M from the single value (0.55) assumed in the previous section to the possible range of D/M, i.e.\ $0.1 \leq \text{D/M} \leq 1$. We do not go to even lower D/M values because our methodology relies on the existence of certain amount of dust.
We derive \aco and fit the \aco-to-\Sigmastar power-law relation at each assumed D/M. The results are shown in \autoref{fig:fit_evolution_DM}.
We highlight the fit results with $0.4 \leq \text{D/M} \leq 0.7$, which is the D/M value inferred from literature introduced in \autoref{sec:calc_aco}. Same as in the previous calculations, the uncertainties of parameters $a$ and $b$ in the fitting parameters are estimated from 1000 rounds of bootstrap resampling.

As shown in the top panel of \autoref{fig:fit_evolution_DM}, we find that the power-law index ($a$) is invariant with assumed D/M throughout the range we examine for both \acotwo and \acoone. The average $a$ in the range of $0.4\leq \mathrm{D/M} \leq 0.7$ is $-0.48^{+0.08}_{-0.09}$ and $-0.22^{+0.08}_{-0.09}$ for \acotwo and \acoone, respectively. The statistical uncertainties in $a$ for both CO transitions are around $\pm0.1$~dex. The result implies that as long as the D/M stays roughly constant within each galaxy, we can recover similar behavior in the \Sigmastar dependence of \aco with a power law

Due to the nature of the definition of $b$ in \autoref{eq:power law}, we expect $b \sim 0$. This is seen in most D/M values we examine as $|b|$ stay below 0.03, shown in the middle panel of \autoref{fig:fit_evolution_DM}. This indicates that the power law parameterization reasonably fits the observed data regardless of the assumed D/M value. However, the $b$ values for \acotwo seems biased toward the positive end. This could result from a steeper $\log$\aco-to-$\log$\Sigmastar slope toward higher \Sigmastar, which results in a positive offset in the power law at relatively lower \Sigmastar.
% \rms / R^2
In the bottom panel, we show the \rms value of each fit as an indicator of goodness of fit. All fits have \rms below 0.2~dex, and the fits around $0.4\leq \mathrm{D/M} \leq 0.7$ have $\rms\sim0.14$~dex.

\begin{figure}
    \centering
	\includegraphics[width=0.99\columnwidth]{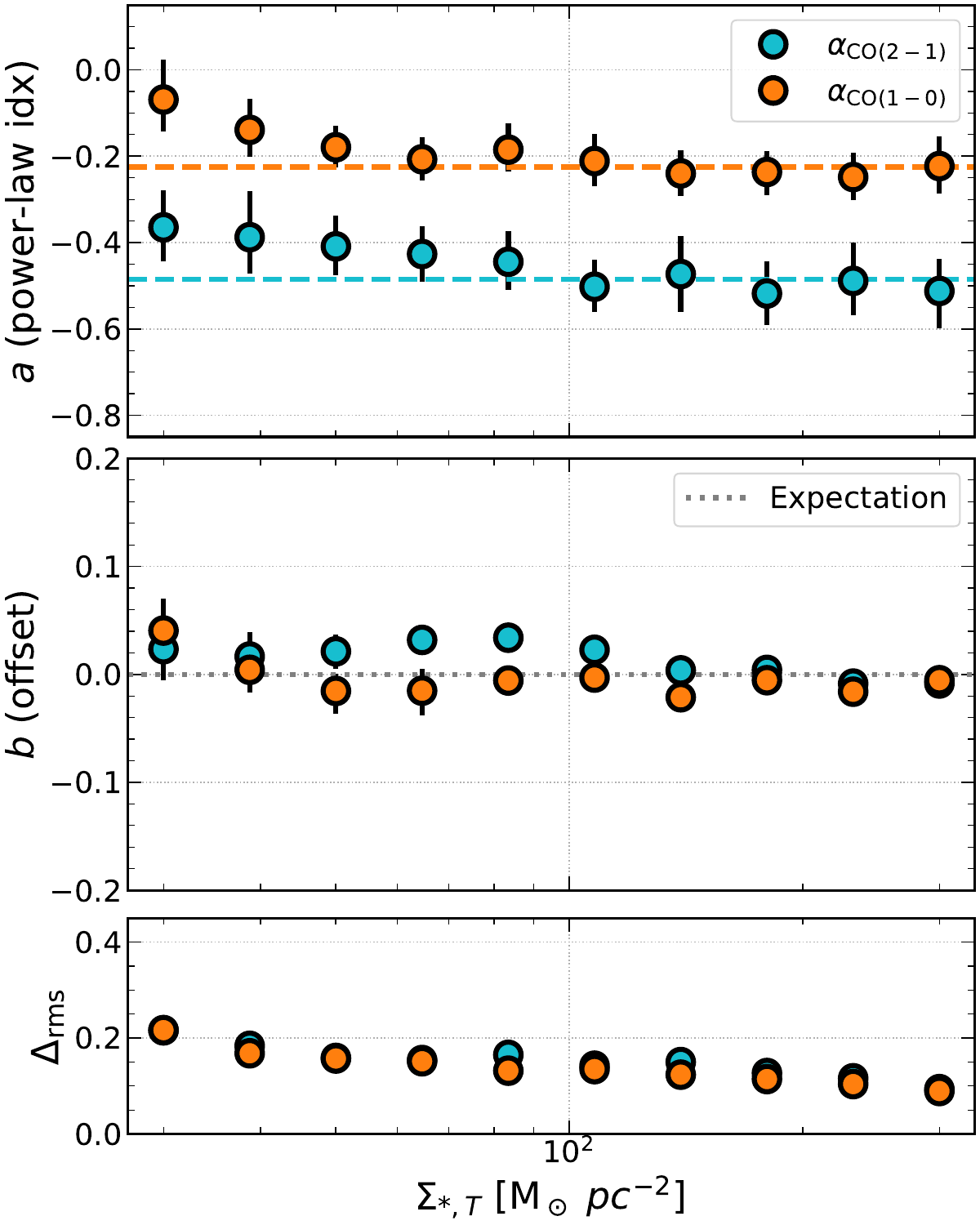}
    \caption{The dependence of fitted power-law index and offset on adopted threshold \SigmastarT. 
    The cyan points show results for \acotwo, and the orange points show results for \acoone. For \acoone, we only show results from \wcoone data only since fitting results including \wcotwo data have mininal difference.
    Top: the power-law index ($a$). The dashed lines show the average $a$ values with $\SigmastarT = 100~\SigmaMassUnit$ in the range of $0.4\leq \mathrm{D/M} \leq 0.7$. These are the same lines as \autoref{fig:fit_evolution_DM}.
    Middle: the offset ($b$). The dotted line shows $b=0$, where expect the fitting result to be if \aco monotonically decreases with \Sigmastar.
    Bottom: the \rms of each fit.
    All the calculation here are done with $\text{D/M}=0.55$.}
    \label{fig:fit_evolution_SST}
\end{figure}

We further test if the chosen threshold in stellar mass surface density, \SigmastarT, will affect the fitting results. We fix $\text{D/M}=0.55$ and fit the power-law relation at \SigmastarT ranges from 30 to 300~\SigmaMassUnit. Note that the number of galaxies included in the subsample changes in each case due to the threshold in \Sigmastar. The results are shown in \autoref{fig:fit_evolution_SST}.

In the top panel of \autoref{fig:fit_evolution_SST}, we notice that the power-law index ($a$) has a larger dynamic range than the case where we alter D/M, but the index stays negative throughout the \SigmastarT range we examine. The power-law index for \acotwo stays within $\pm0.1$ of the fiducial case, and the indices for \acoone is consistent with the fiducial value at $\SigmastarT>60~\SigmaMassUnit$. There is a weak trend that $|a|$ becomes larger toward larger \SigmastarT. 
The small $b$ values indicate that the power-law function form applies in general.
We also show the \rms values in the bottom panel. We have $\rms \leq 0.2$, with a weak trend of smaller \rms toward higher \SigmastarT.

To summarize, the power-law functional form applies to the normalized \aco-to-\Sigmastar relation within the D/M and \SigmastarT ranges we examine.
With a fixed \SigmastarT at 100~\SigmaMassUnit, we find a invariant power-law index ($a$) throughout $0.1 \leq \mathrm{D/M} < 1.0$. The average $a$ in the range of $0.4\leq \mathrm{D/M} \leq 0.7$ is $-0.48^{+0.08}_{-0.09}$ and $-0.22^{+0.08}_{-0.09}$ for \acotwo and \acoone, respectively.
With a fixed D/M at 0.55, we find a weak trend of larger $|a|$ toward higher \SigmastarT.
The power-law index derived with the fiducial setup, i.e.\ $\mathrm{D/M}=0.55$ and $\SigmastarT=100~\SigmaMassUnit$, is a good representative of conditions with $60 < \SigmastarT \leq 300~\SigmaMassUnit$, with a span $\sim\pm0.09$.

\subsection{Galaxy-to-galaxy variations}\label{sec:aCO-SigmaMstar:galaxy}
\begin{figure*}
    \centering
	\includegraphics[width=0.99\textwidth]{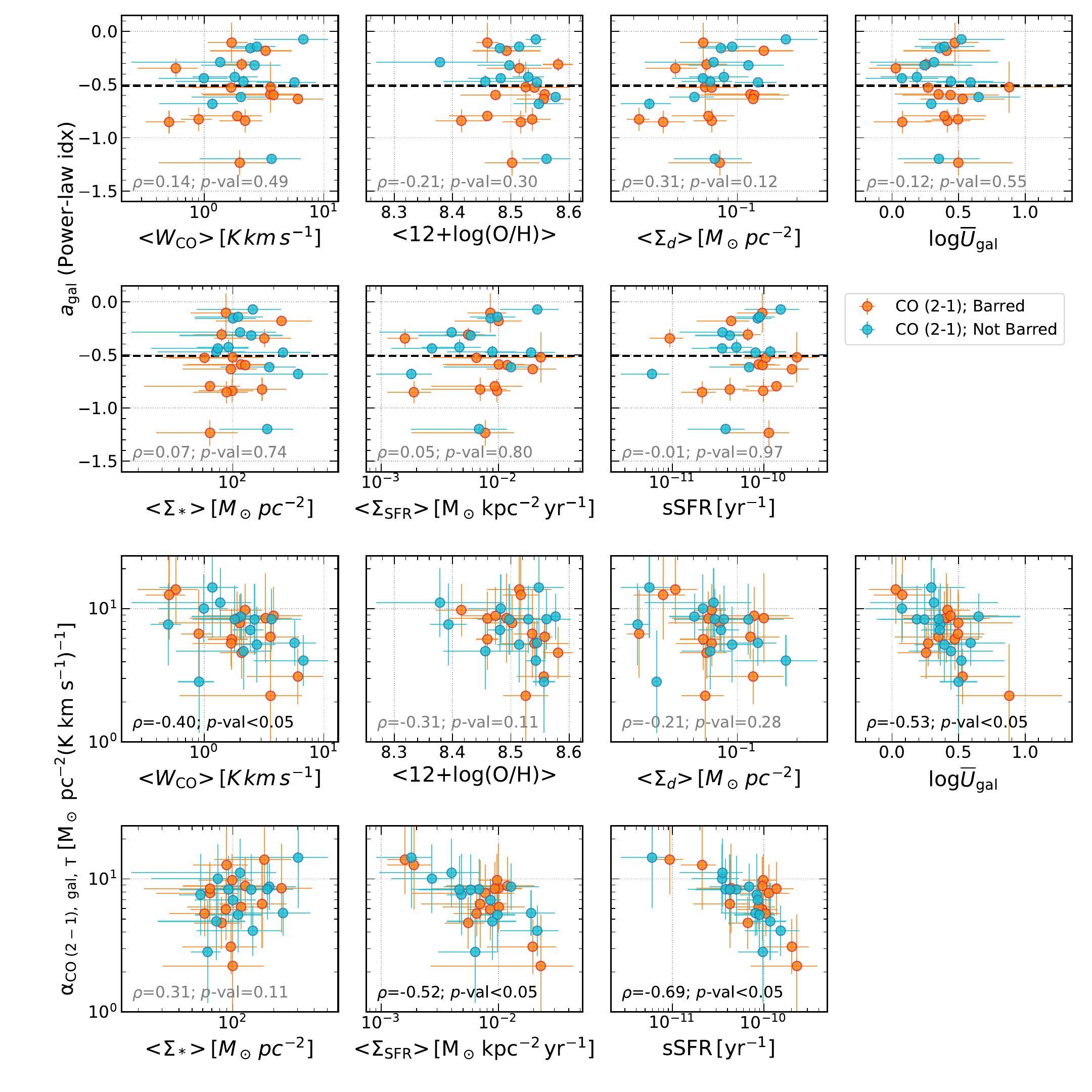}
    \caption{Scaling relations between $a_\mathrm{gal}$ (the power-law index), \acoT and galaxy-averaged environmental parameters for \cotwo data.
    Galaxies that are barred and not barred are colored in orange and cyan, respectively.
    The black dashed line in the $a_\mathrm{gal}$ show the $a$ value calculated with overall data.
    The correlation coefficients and $p$-values are labeled at the lower left in each panel, highlighting the significant ($p$-value$<0.05$) ones.}
    \label{fig:plidx_norm_21}
\end{figure*}

\begin{figure*}
    \centering
	\includegraphics[width=0.99\textwidth]{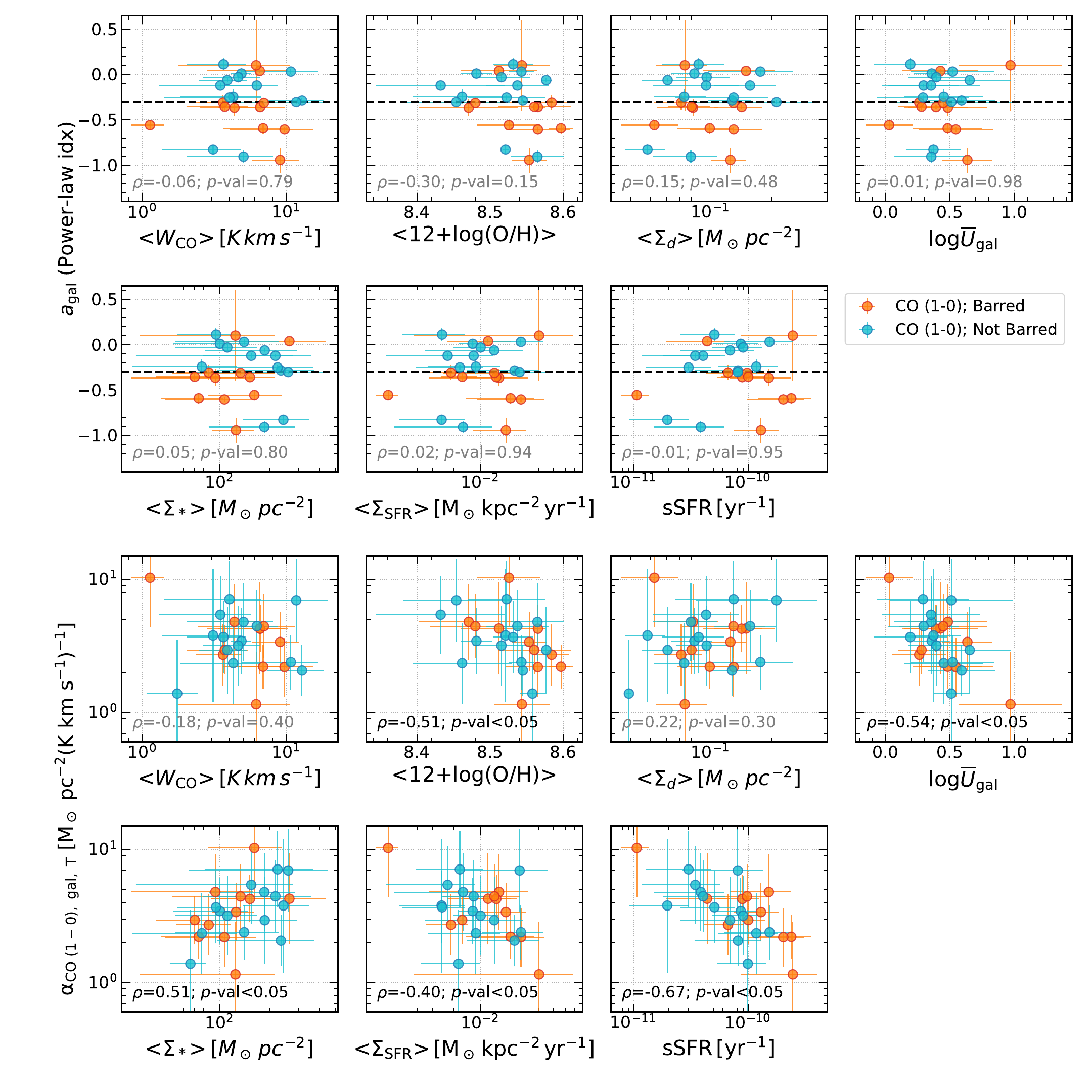}
    \caption{Similar to \autoref{fig:plidx_norm_21}, but for \coone data.}
    \label{fig:plidx_norm_10}
\end{figure*}

In this section, we examine the possible variation of the \aco-to-\Sigmastar relation between individual galaxies, mainly how the variation in \acoT (normalization of \aco at \SigmastarT, see \autoref{sec:aCO-SigmaMstar}) and $a_\mathrm{gal}$ (power-law index, \autoref{eq:power law}) correlate with galaxy-averaged properties. 
By understanding what sets $a_\mathrm{gal}$ and \acoT, we can build a prescription of \aco considering the \aco-to-\Sigmastar relation and galaxy-to-galaxy variations.
The results are visualized in \autoref{fig:plidx_norm_21} and \autoref{fig:plidx_norm_10}.

In the upper panels of \autoref{fig:plidx_norm_21} and \autoref{fig:plidx_norm_10}, we show how the power-law index ($a_\mathrm{gal}$) varies with 7 selected galaxy-averaged properties and whether the galaxy is barred or not. The set of properties is the same as the ones in \autoref{fig:mean_aCO21_properties} and \autoref{fig:mean_aCO10_properties}. None of the properties show significant correlation with $a_\mathrm{gal}$. Meanwhile, the standard deviation of $a_\mathrm{gal}$ is 0.30 for both \cotwo and \coone.

In the lower panels of \autoref{fig:plidx_norm_21} and \autoref{fig:plidx_norm_10}, we show how the normalization in each galaxy (\acoT) varies with galaxy-averaged properties. The standard deviation of \acoT is 0.2~dex for both \cotwo and \coone. For \cotwo, $\text{<}W_\mathrm{CO}\text{>}$, $\log\Ubar_\mathrm{gal}$, $\text{<}\Sigmasfr\text{>}$, and sSFR show significant correlations with \acoT with similar strength. For \coone, $\text{<}\metal\text{>}$, $\log\Ubar_\mathrm{gal}$, $\text{<}\Sigmastar\text{>}$, $\text{<}\Sigmasfr\text{>}$ and sSFR show significant correlations with \acoT with similar strength. We use these significant correlations to fit empirical relations for \acoT and summarize the results in \autoref{tab: aCOT fits}. The fitted empirical relations do not differ significantly in terms of \rms. Meanwhile, the fit with \Ubar has the smallest\footnote{Considering the product of $\log x$ and the uncertainties in $m$.} the statistical uncertainties in the fitted parameters among the parameters for both \cotwo and \coone. Besides \Ubar, <$\Sigma_\mathrm{SFR}$> also has small statistical uncertainties and is available for more galaxies.

\begin{center}
\begin{table}
\caption{Dependence of \acoT on galaxy-integrated quantities.}
\label{tab: aCOT fits}
\begin{tabular}{llll}
\toprule
\multicolumn{4}{c}{$\log\alpha_{\rm \cotwo,gal,T}$, \cotwo Only}\\
\midrule
$x^{(1)}$ & $m^{(1)}$ & $d^{(1)}$ & \rms \\
\midrule
<$W_{\rm CO}$> & -0.4$\pm$0.1 & 1.0$\pm$0.1 & 0.15 \\
$\overline{U}_{\rm gal}$ & -0.4$\pm$0.2 & 1.0$\pm$0.1 & 0.16 \\
<$\Sigma_{\rm SFR}$> & -0.4$\pm$0.1 & 0.0$\pm$0.2 & 0.13 \\
sSFR & -0.5$\pm$0.1 & -3.9$\pm$0.9 & 0.11 \\
\toprule
\multicolumn{4}{c}{$\log\alpha_{\rm \coone,gal,T}$, \coone Only}\\
\midrule
$x$ & $m$ & $d$ & \rms \\
\midrule
<12+log(O/H)> & -2.0$\pm$0.9 & 17.8$\pm$7.4 & 0.17 \\
$\overline{U}_{\rm gal}$ & -0.7$\pm$0.3 & 0.8$\pm$0.1 & 0.16 \\
<$\Sigma_\star$> & 0.5$\pm$0.2 & -0.4$\pm$0.4 & 0.17 \\
<$\Sigma_{\rm SFR}$> & -0.4$\pm$0.2 & -0.2$\pm$0.3 & 0.16 \\
sSFR & -0.4$\pm$0.1 & -3.9$\pm$1.0 & 0.13 \\
\bottomrule
\end{tabular}
\tablecomments{(1) The linear regression formula is $\log\acoT=m\log x+d$ for most quantities and $\log\acoT=mx+d$ for <12+log(O/H)>}. $W_\mathrm{CO}$ in [\ICOUnit], \Sigmasfr in [\SigmasfrUnit], \Sigmastar in [\SigmaMassUnit] and sSFR in [yr$^{-1}$].
\end{table}
\end{center}

\section{Discussion}\label{sec:discussions}

\subsection{General suggestions for \texorpdfstring{\aco}{conversion factor} prescriptions}\label{sec:aCO-SigmaMstar:suggestion}

In \autoref{sec:results:local}, we present how the measured \aco correlates with local physical quantities and provide linear regression for each quantity at 2~kpc scale in \autoref{tab:aco_stats}. These measurements consider the statistical behavior of the overall sample. Among the quantities, $W_\mathrm{CO}$, \Ubar, and \Sigmasfr usually have the strongest correlations with \aco and smallest \rms from linear regression. We would suggest the readers go with these parameters if the parameter space of their sample overlaps with this study (see \autoref{fig:aCO10_completeness} for the completeness of each quantity).

Meanwhile, we explicitly explore the relation between \aco and \Sigmastar in \autoref{sec:aCO-SigmaMstar} as a possible tracer for starburst \aco at 2~kpc scale, and find strong correlation between \aco and \Sigmastar after normalization at some \Sigmastar threshold. There are two ways to adopt these results. The first one is a stand-alone prescription combining the indices from \autoref{sec:aCO-SigmaMstar:D/M} and normalization from \autoref{tab: aCOT fits}, using galaxy-averaged $\overline{U}_{\rm gal}$ as an example:
\begin{equation}
\left\{
\begin{array}{l}
    \acotwo = 
    10^{-0.4\log(\overline{U}_{\rm gal})+1.0}\Sigmastar^{-0.48}\\
    \acoone = 
    10^{-0.7\log(\overline{U}_{\rm gal})+0.8}\Sigmastar^{-0.22}
\end{array}
,~\Sigmastar\geq 100~\SigmaMassUnit,\right.
\end{equation}
where \aco is in unit of \acoUnit, and the $\overline{U}_{\rm gal}$-dependent normalization could be replaced with other quantities listed in \autoref{tab: aCOT fits}, e.g.\ <$\Sigma_\mathrm{SFR}$>. Please refer to \autoref{sec:aCO-SigmaMstar} for relevant uncertainties. We note that the possible variation of the power-law index could be up to $\sim 0.2$ due to internal variations of D/M (\autoref{app:d/m}).

On the other hand, one key mechanism that sets \aco, the CO-dark gas, is likely not parameterized by our formula (see \autoref{sec:discussions:interpret}). This is because the CO-dark gas effect is relatively weak in the metallicity span of our sample. However, both the CO-dark gas and the ``starburst \aco'' effects should be considered for an \aco prescription to be applied through all environments. Thus, another suggestion we have is to make a \citet{BOLATTO13}-style combination (also see E. Schinnerer \& A. K. Leroy ARA\&A submitted) of our \Sigmastar power-law term with existing \aco prescriptions tracing the CO-dark gas effect, e.g.\ \citet{WOLFIRE10}, \citet{NARAYANAN12}, \citet{SCHRUBA12}, \citet{HUNT15}, \citet{ACCURSO17}, or \citet{Sun20}. That is, assuming the adopted existing CO-dark prescription is $\aco^\mathrm{CO\text{-}dark}$, we suggest:
\begin{equation}
    \acotwo = \left\{\begin{array}{ll}
    \aco^\mathrm{CO\text{-}dark} & ,~\Sigmastar< 60~\SigmaMassUnit \\
    \aco^\mathrm{CO\text{-}dark}\Sigmastar^{-0.48} & ,~\Sigmastar\geq 60~\SigmaMassUnit
    \end{array}\right..
\end{equation}
Under this functional form, we expect the normalization (\acoT) to be taken into account by the $\aco^\mathrm{CO\text{-}dark}$ term. Since this formula does not include our own normalization, the \Sigmastar power-law can be extended to lower \Sigmastar threshold. For \acoone case, one can simply replace the power-law index with $-0.22$.

One of the future directions we will take is to study how \aco correlates with physical quantities at cloud scales instead of kpc scales and build \aco prescriptions accordingly.
The advantage in this direction is that the physical quantities at cloud scale are more strongly linked to fundamental physics of CO emission and dynamics of molecular clouds.
For instance, based on the \aco measurements in this work, \citet{Teng24_alphaCO_dv} have reported \aco dependence with cloud-scale velocity dispersion which likely traces CO opacity change. We also refer the readers to \citet{Teng22,Teng23} for more details.

The other possible future direction for dust-based \aco is a new strategy that simultaneously allows variations in \aco, dust properties (e.g.\ D/M in \autoref{app:d/m} and dust opacity in \autoref{sec:data:data sets}) and metallicity at best-possible resolution. The \citet{LEROY11} and \citet[][also see their Appendix A]{SANDSTROM13} strategy is a good demonstration of the concept for most items on the list except the resolution. A more sophisticated strategy would help identify the next step forward on \aco prescriptions. We are also interested in investigating whether the \Sigmastar-dependence still applies to \Sigmagas-dominated environments.

\subsection{Interpreting of the environmental dependence of \texorpdfstring{\aco}{}}\label{sec:discussions:interpret}
In this section, we will discuss the physical interpretations of the correlations between \aco and the physical quantities we present in the previous sections. As we mentioned in \autoref{sec:intro}, we expect two main trends in the variation of \aco: (1) the ``CO-dark gas'' trend, where \aco increases toward lower metallicity as shielding for CO weakens; (2) the ``starburst conversion factor'' trend, where \aco decreases toward galaxy centers and U/LIRGs with the decrease in CO optical depth or increase in CO excitation.

Regarding the CO-dark gas trend, we observe moderate to weak anti-correlation between \aco and \metal at kpc scale (\autoref{sec:results:local}), and moderate anti-correlation at galaxy scale (\autoref{sec:results:global}). One possible explanation for the weak correlation is that the statistical significance becomes weaker with the small dynamic range of our \metal data: 80\% of the \metal measurements fall within a 0.2~dex range from 8.4 to 8.6. The \aco-\Sigmastar relation we present in \autoref{sec:aCO-SigmaMstar} is unlikely caused by this CO-dark gas effect since the dynamic range in \metal is even smaller for data above the \SigmastarT threshold. Another explanation is that the CO-dark gas effect is weaker at nearly solar metallicity \citep[e.g.][]{WOLFIRE10,GloverMacLow11,HUNT15}. However, we note that recent simulations show that there is significant fraction of CO-dark gas ($f_\mathrm{dark}$) up to solar metallicity, e.g.\ \citet{Gong18} found $f_\mathrm{dark}$ ranges from 26\%--79\%. These studies found that $f_\mathrm{dark}$ correlates with extinction and/or $W_\mathrm{CO}$ \citep{SMITH14,Gong18,Gong20,Hu22}.

We interpret \Ubar and (r)sSFR as empirical tracers for regions with high SFR, where the ``starburst conversion factor'' trend matters and lowers down \aco. Some studies have also argued that \aco could decrease with increased radiation field due to CO dissociation \citep{ISRAEL97,WOLFIRE10,ACCURSO17}. However, we did not observe this trend, and one possible explanation is that the CO dissociation effect should be weak as long as the CO emission is optically thick \citep{WOLFIRE10,BOLATTO13}. \Sigmasfr, which simultaneously traces the UV radiation and starburst regions, also have moderate anti-correlation with \aco across all cases. In general, we observe a stronger correlation between \Sigmasfr and \acotwo than \acoone. We also observe moderate anti-correlation between $W_\mathrm{CO}$ and \aco. This is consistent with the theoretical assumption under optically thick assumption \citet{BOLATTO13} and recent observations \citep{Hunt23}.

We interpret the \aco-to-\Sigmastar anti-correlation as the increase of velocity dispersion of molecular gas from additional gravity. \citet[][also see \citet{Hirashita23b_aCO}]{BOLATTO13} suggested that in high-\Sigmastar environments, 
the molecular gas experiences gravitational potential from stellar sources, ending up with a total pressure larger than isolated, virialized clouds. This larger pressure results in gas line width larger than one would expect from self-gravitating clouds. This increase in gas line width scales with total mass (stars and gas) in the system, or $\aco \propto \big(M_\mathrm{mol}/(M_\mathrm{mol}+M_\star)\big)^{0.5}$. The above functional form approximates a \aco-to-\Sigmastar power law in $M_\star$-dominated regions. For the argument to hold, the CO emission must be optically thick. \citet{BOLATTO13} mentioned that the only possible structure for molecular gas that satisfies this scenario is an extended molecular medium.

\subsection{Comparing to previous \texorpdfstring{\aco}{} surveys}
% S13
% COMING: Table 3 summary of global measurements

In this section, we compare our measurements to \aco values obtained in previous dust-based \aco surveys. Since we will cover studies with both \acotwo and \acoone measurements, we will convert \acotwo values in literature and this work to \acoone with $R_{21}=0.65$ for simplicity and uniformity.
First, we compare our \aco maps with \citet{SANDSTROM13}. They measured spatially resolved \aco in 26 nearby, star-forming galaxies using a dust-based methodology \citep[also see][]{LEROY11,DenBrok23,Yasuda23}. They assume that the variation of D/G is minimal in a few-kpc-scale ``solution pixel'' consisting of 37 samples in a hexagonal region, and fit D/G and \aco simultaneously from data by minimizing the variation in D/G. \citet{SANDSTROM13} used \cotwo data from HERACLES \citep{LEROY09} and $R_{21}=0.7$. We rescale their results with $R_{21}=0.65$ for uniformity. Compared to our work, the \citet{SANDSTROM13} metholodogy has larger degrees of freedom for the spatial variation of D/G and D/M. Meanwhile, it is more difficult to push their methodology to a larger sample size at fixed physical resolution. 

\begin{figure}
    \centering
	\includegraphics[width=\columnwidth]{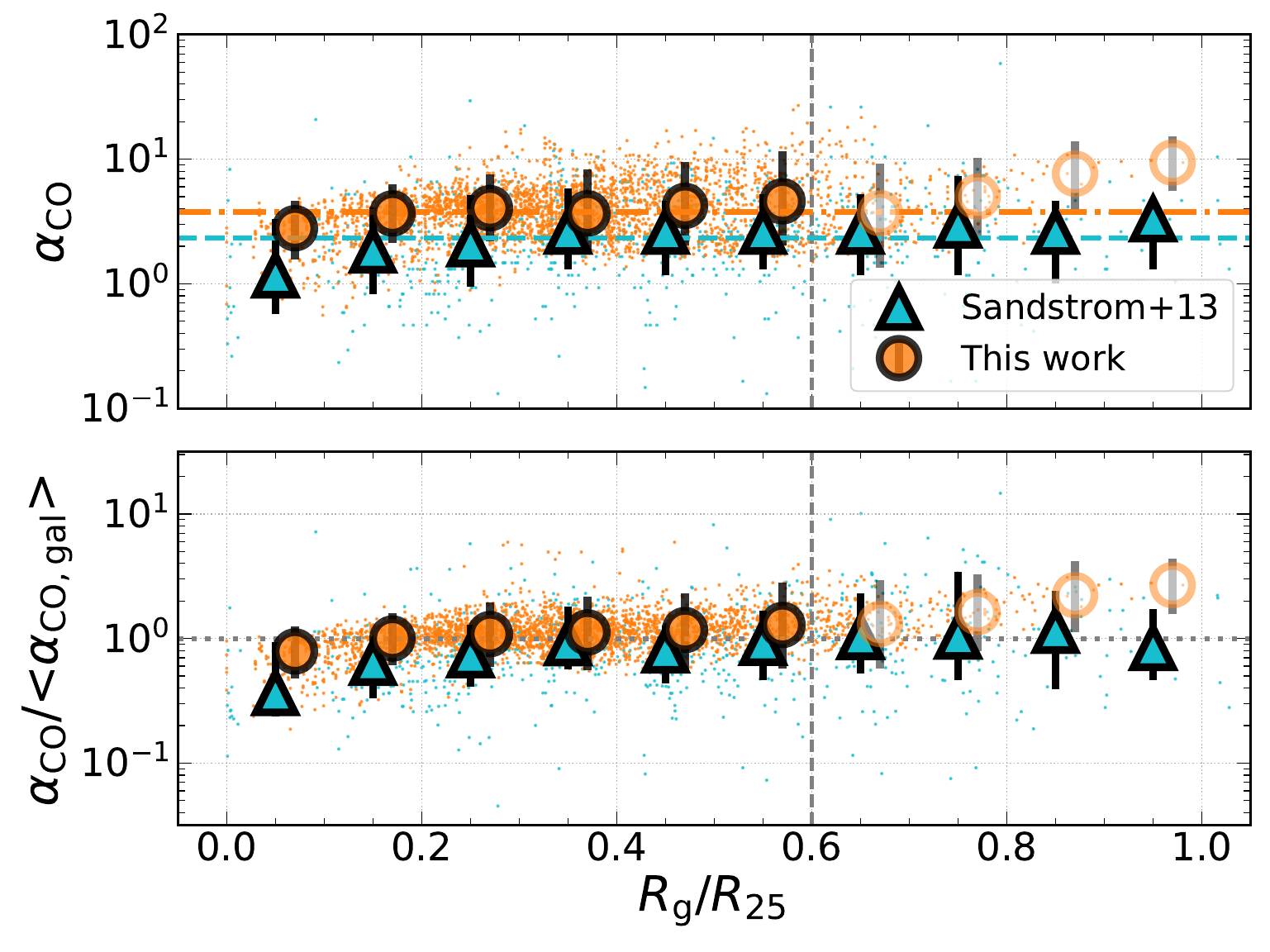}
    \caption{The relation between measured \aco and galacto-centric radius in this work and \citet{SANDSTROM13}. For both works, we display $\aco=0.65\acotwo$ for uniformity. We only include measurements from the galaxies that are included in both works.
    The vertical dashed line shows the completeness threshold in $R_\mathrm{g}/R_{25}$ in this work.
    In the top panel, the horizontal cyan dashed-dotted line shows the mean \aco in \citet{SANDSTROM13}, and the orange dashed line shows the mean \aco from this work. The two lines are close to each other.
    In the bottom panel, the horizontal line shows $\aco=\text{<}\alpha_\mathrm{CO,~gal}\text{>}$, where    <$\alpha_\mathrm{CO,~gal}$> is the mean \aco in each galaxy.
    }
    \label{fig:aCO_S13}
\end{figure}

There are 13 galaxies that are studied in both \citet[][see their Figure 7]{SANDSTROM13} and this work.
% : NCG0628, NGC2841, NGC3184, NGC3351, NGC3521, NGC3627, NGC3938, NGC4254, NGC4321, NGC4536, NGC4569, NGC4725, NGC4736, NGC5055, NGC5457, NGC6946, and NGC7331. 
We show the \aco measurements from both works as a function of galacto-centric radius ($R_\mathrm{g}$ in terms of $R_{25}$.) in \autoref{fig:aCO_S13}.  We adopt $R_{25}$ values in this work instead of the \citet{SANDSTROM13} values. The simple mean \aco of all measurements in \citet{SANDSTROM13} is $\sim 2.3~\acoUnit$. Similar with \citet{SANDSTROM13}, we find a weak to moderate (but significant) positive correlation between \aco and $R_\mathrm{g}$.
When we normalize the \aco in each galaxy by the mean \aco in each galaxy (<$\alpha_\mathrm{CO,~gal}$>), both works show a flat trend with radius in mid- to outer-disk. \citet{SANDSTROM13} data shows a more significant decrease in \aco in the galaxy center. Several factors could contribute to the difference in the inner most radial bins. If we calculate the $W_\mathrm{CO}$-weighted mean instead of the median in each bin, the difference in the bin with smallest radii will decrease by 0.1~dex, which partially explains the discrepancy. Another possible explanation is that some of the measurements with small \Sigmamol (and possibly small \aco) are removed due to small S/N; however, they are taken into account in \citet{SANDSTROM13}. It is not clear whether the difference in resolution is a cause. When we calculate \aco at 1~kpc resolution with the galaxies with distance within 10~Mpc, there is no clear trend of the resulting \aco with resolution. The fixed D/M is unlikely to be a major cause since the \citet{SANDSTROM13} results are consistent with a D/G-metallicity power law, see their Figure~13.

\begin{figure*}
    \centering
	\includegraphics[width=\textwidth]{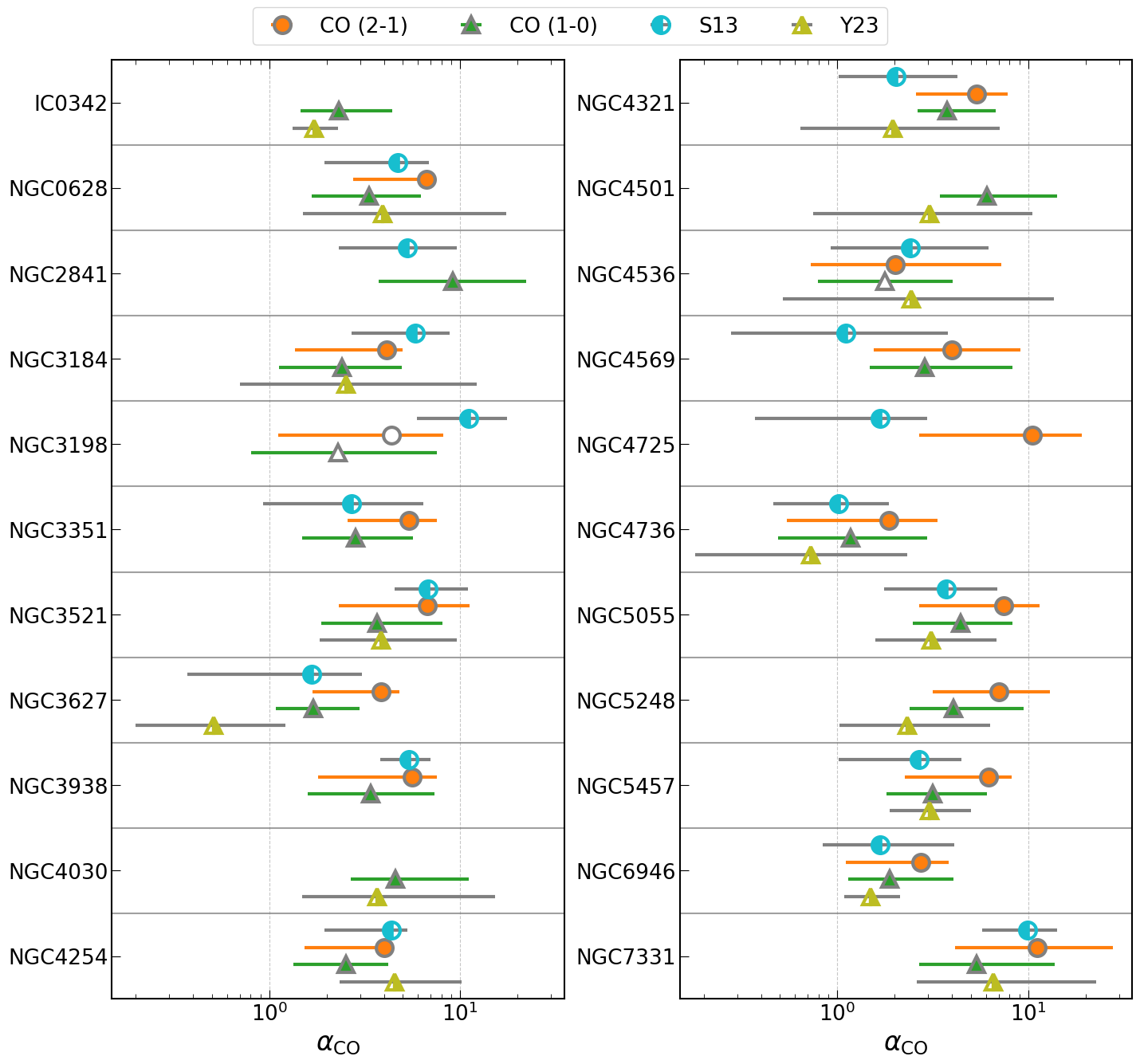}
    \caption{The mean \aco values in each galaxy from this work (both \cotwo data with $R_{21}=0.7$ and \coone data), \citet[][S13]{SANDSTROM13} and the COMING survey \citep[][Y23]{Sorai19_COMING,Yasuda23}. 
    Circles show \aco values derived with \cotwo data, and triangles show \aco values derived with \coone data. Filled symbols show the results from this work, empty symbols show the ones with low CO recovery fraction (\autoref{tab:aco_stats}), and half-filled symbols show literature values.
    We only include galaxies that are measured in at least one of the literature survey. The errorbar for this work shows the 16th- and 84th-percentiles (\autoref{tab:aco_stats}). The mean and errorbar of previous works are quoted from Table 4 of \citet[][with rescaling for $R_{21}$]{SANDSTROM13} and Table 3 of \citet{Yasuda23}. The mean values are $W_\mathrm{CO}$-weighted mean for this work and \citet{SANDSTROM13}, and the ``global'' result for \citet{Yasuda23}.
    }
    \label{fig:previous_works}
\end{figure*}

For comparing galaxy-averaged \aco measurements, we include another previous study: the COMING survey \citep{Sorai19_COMING,Yasuda23}. The COMING survey solves D/G and \acoone simultaneously by minimizing a $\chi^2$ value defined by the difference between $\big(\text{D/G}\times(\Sigmaatom+\acoone\wcoone)\big)$ and \Sigmad derived from dust SED fitting. Here, we quote their ``global'' result, where the authors fit all data within one galaxy to retrieve one set of D/G and \aco values.

We compare our measured \aco in each galaxy with \citet{SANDSTROM13} and the COMING survey \citep{Sorai19_COMING,Yasuda23} in \autoref{fig:previous_works}. For \citet{SANDSTROM13} and this work, we adopt the $W_\mathrm{CO}$-weighted mean. For the COMING survey, we adopt their ``global'' result. The \aco measured in the three works are in general consistent with each other within uncertainties.
Our measurements made with \cotwo and \coone agree with each other. When there is a difference, it is more often that the one derived with \cotwo has a slightly larger value.
% There are a few cases where our measured \aco is significantly lower than the one in \citet{SANDSTROM13}, i.e.\ NGC0925, NGC3184, NGC3198, and NGC3938. Since the two works use almost identical data sets for gas emission lines in these targets, we believe the main difference results from the estimate of \Sigmad. In this work, we apply the correction factor for \Sigmad derived in \citet{Chastenet21_M101} for dust properties fitted with the \citet{DRAINE07} model. This can cause a $\sim 0.5$~dex lower \aco than the one directly derived from the \citet{DRAINE07} model.
We also note that in several galaxies with signatures of active galactic nucleus \citep[AGN; see classification in][]{KENNICUTT11(KINGFISH)2011PASP..123.1347K}, there is larger offset between our measurements and literature, e.g.\ NGC3627 and NGC4725; however, there are also galaxies with AGN show consistent results, e.g.\ NGC4536, NGC4736 and NGC5055. Thus, having AGN is not the only cause for the mismatch, and it is likely that the type of nuclei activities do not dominate the kpc-scale \aco values \citep[e.g.][]{SANDSTROM13}. The adopted dust SED fitting method is also unlikely the cause for the difference in NGC4725 since we have a lower estimate of \Sigmad, which should yield smaller \aco. 
Our measurements made with \coone generally agree with the COMING survey.

\section{Summary}\label{sec:summary}

In this work, we measure the spatially resolved CO-to-H$_2$ conversion factor (\aco) in \nsample nearby galaxies at 2~kpc resolution. We derive \Sigmamol by using a fixed D/M and converting \Sigmad and $Z$ into \Sigmagas, then removing \Sigmaatom to get \Sigmamol. We calculate \aco with derived \Sigmamol and measured $W_\mathrm{CO}$. In total, we have $\sim$810 and $\sim$610 independent measurements of \aco for \cotwo and \coone data, respectively. The mean values for \acotwo and \acoone are $9.7^{+4.7}_{-5.7}$ and $4.2^{+1.9}_{-2.0}~\acoUnit$, respectively. The CO-intensity-weighted mean for \acotwo is 5.76~\acoUnit, and 3.33~\acoUnit for \acoone. These values are measured in 37 galaxies with data $\mathrm{S/N} > 1$.

We examine how \aco scales with several physical quantities, i.e.\ $W_\mathrm{CO}$, metallicity, \Sigmad, ISRF, \Sigmastar, \Sigmasfr and (r)sSFR. At 2~kpc scale, all quantities have significant local correlation with \aco. Among them, the strength of the ISRF (\Ubar), \Sigmasfr and $W_\mathrm{CO}$ have the strongest anti-correlation with spatially resolved $\alpha_{\rm CO}$. We provide linear regression of \aco with all the quantities tested, along with the corresponding performance and uncertainties in \autoref{tab: aCO_prop}.

At galaxy-integrated scale, most quantities have significant correlation with $W_\mathrm{CO}$-weighted mean \aco. \Ubar, \Sigmasfr, $W_\mathrm{CO}$ and \metal have significant correlations with \aco for both \coone and \cotwo cases.

When we normalized resolved \aco measurements by the $W_\mathrm{CO}$-weighted mean in each galaxy, we find an increased correlation strength between normalized \aco and \Sigmastar. After examining through \Sigmastar bins, we find that in regions with high stellar mass surface densities ($\Sigmastar\geq 100~\SigmaMassUnit$), the \aco decreases with \Sigmastar. Specifically, we find:
\begin{equation}
    \left\{
    \begin{array}{l}
         \acotwo \propto \Sigmastar^{-0.48}  \\
         \acoone \propto \Sigmastar^{-0.22}
    \end{array}
    ,~\Sigmastar\geq 100~\SigmaMassUnit\right.
\end{equation}
within D/M=0.4--0.7, the D/M values for inner disk inferred from literature. The power-law index is insensitive to the assumed D/M, and it is roughly constant at $\SigmastarT> 60~\SigmaMassUnit$. It also has little dependence on the adopted ratio between CO rotational lines. % Meanwhile, the discrepancy between the power-law index for \cotwo and \coone data indicates a variation in the CO line ratio traced by \Sigmastar.

When fitting the power-law relation within individual galaxies, we find significant dependence of the normalization of \aco in each galaxy on several quantities. Among them, the linear regression to $\log\Ubar_\mathrm{gal}$ has the minimal statistical uncertainties. Thus, we recommend using \Sigmastar and $\log\Ubar_\mathrm{gal}$ to predict \aco at high-\Sigmastar environments.

This decrease of \aco in the high-\Sigmastar region is likely due to the increased CO brightness with increased line width. The line width is larger than self-gravitating clouds due to the additional gravity from stellar sources, and the structure satisfying this scenario is likely an extended molecular medium. Understanding the decrease in \aco at high \Sigmastar is important for accurately assessing molecular gas content and star-formation efficiency in the centers of galaxies and bridges the ``MW-like'' to "starburst" conversion factor.

\section*{Acknowledgments}
IC thanks Hiroyuki Hirashita for helpful discussion.
IC thanks the National Science and Technology Council for support through grant 111-2112-M-001-038-MY3, and the Academia Sinica for Investigator Award AS-IA-109-M02 (PI: Hiroyuki Hirashita).
KS and YT acknowledge funding support from NRAO Student Observing Support Grant SOSPADA-012 and from the National Science Foundation (NSF) under grant No. 2108081.
JS acknowledges support by NASA through the NASA Hubble Fellowship grant HST-HF2-51544 awarded by the Space Telescope Science Institute (STScI), which is operated by the Association of Universities for Research in Astronomy, Inc., for NASA, under contract NAS5-26555. ADB acknowledges support from the National Science Foundation through grants AST-2108140 and AST-2307441.
EWK acknowledges support from the Smithsonian Institution as a Submillimeter Array (SMA) Fellow and the Natural Sciences and Engineering Research Council of Canada.
JC acknowledges support from ERC starting grant \#851622 DustOrigin
% \textcolor{blue}{Please add your acknowledgements here.}

This work uses observations made with ESA \textit{Herschel} Space Observatory. \textit{Herschel} is an ESA space observatory with science instruments provided by European-led Principal Investigator consortia and with important participation from NASA. The \textit{Herschel} spacecraft was designed, built, tested, and launched under a contract to ESA managed by the \textit{Herschel}/Planck Project team by an industrial consortium under the overall responsibility of the prime contractor Thales Alenia Space (Cannes), and including Astrium (Friedrichshafen) responsible for the payload module and for system testing at spacecraft level, Thales Alenia Space (Turin) responsible for the service module, and Astrium (Toulouse) responsible for the telescope, with in excess of a hundred subcontractors.

This paper makes use of the VLA data with project codes 14A-468, 14B-396, 16A-275 and 17A-073, which has been processed as part of the EveryTHINGS survey.
This paper makes use of the VLA data with legacy ID AU157, which has been processed in the PHANGS--VLA survey.
The National Radio Astronomy Observatory is a facility of the National Science Foundation operated under cooperative agreement by Associated Universities, Inc. 
This publication makes use of data products from the Wide-field Infrared Survey Explorer, which is a joint project of the University of California, Los Angeles, and the Jet Propulsion Laboratory/California Institute of Technology, funded by the National Aeronautics and Space Administration.

This paper makes use of the following ALMA data, which have been processed as part of the PHANGS--ALMA CO(2--1) survey: \linebreak
ADS/JAO.ALMA\#2012.1.00650.S, \linebreak % (N628/M74)
% ADS/JAO.ALMA\#2013.1.00803.S, \linebreak % (N5128/CenA)
%ADS/JAO.ALMA\#2013.1.01161.S, \linebreak % (N1365 + N5236/M83)
% ADS/JAO.ALMA\#2015.1.00121.S, \linebreak % (N5236/M83)
ADS/JAO.ALMA\#2015.1.00782.S, \linebreak % (N1313 + N7793)
%ADS/JAO.ALMA\#2015.1.00925.S, \linebreak % (pilot low mass)
%ADS/JAO.ALMA\#2015.1.00956.S, \linebreak % (pilot high mass)
% ADS/JAO.ALMA\#2016.1.00386.S, \linebreak % (N5236/M83)
%ADS/JAO.ALMA\#2017.1.00392.S, \linebreak % (low mass follow-up)
% ADS/JAO.ALMA\#2017.1.00766.S, \linebreak % (early-type)
%ADS/JAO.ALMA\#2017.1.00886.L, \linebreak % (large program)
ADS/JAO.ALMA\#2018.1.01321.S, \linebreak % (N253, N300, Circinus)
ADS/JAO.ALMA\#2018.1.01651.S. \linebreak % (main sample follow-up)
% ADS/JAO.ALMA\#2018.A.00062.S. \linebreak % (ACA-only nearby)
ALMA is a partnership of ESO (representing its member states), NSF (USA) and NINS (Japan), together with NRC (Canada), MOST and ASIAA (Taiwan), and KASI (Republic of Korea), in cooperation with the Republic of Chile. The Joint ALMA Observatory is operated by ESO, AUI/NRAO and NAOJ.

This research made use of Astropy,\footnote{http://www.astropy.org} a community-developed core Python package for Astronomy \citep{ASTROPY13,Astropy18,Astropy22}. 
This research has made use of NASA's Astrophysics Data System Bibliographic Services. 
We acknowledge the usage of the HyperLeda database (http://leda.univ-lyon1.fr). 
This research has made use of the NASA/IPAC Extragalactic Database (NED), which is funded by the National Aeronautics and Space Administration and operated by the California Institute of Technology.

\vspace{5mm}
% Facility keywords from https://journals.aas.org/facility-keywords/
\facilities{ALMA, \textit{Herschel}, IRAM:30m, VLA, WISE, WSRT}
\software{
    astropy \citep{ASTROPY13,Astropy18,Astropy22},
    matplotlib \citep{HUNTER07},
    numpy \& scipy \citep{VANDERWALT11},
    }

\appendix

\section{Internal variation of D/M}\label{app:d/m}

It is questionable whether D/M is a constant within galaxies, even in galaxy centers. Observations have found internal variations of D/M within galaxies \citep{JENKINS09,ROMAN-DUVAL14,ROMAN-DUVAL17,CHIANG18,Vilchez19}. In these studies, people found a higher D/M toward higher metallicity or gas surface densities. A varying D/M within galaxies is also expected by several models \citep{HOU19,LI19,AOYAMA20_2020MNRAS.491.3844A}. However, how to characterize the variation of D/M with local condition is a topic remained unsolved and it is outside the main scope of this work.

To demonstrate how a varying D/M might affect our results, we define a toy model with D/M increasing toward galaxy centers. For the galaxy disks ($R_\mathrm{g} > R_\mathrm{e}$), we assume $\mathrm{D/M}=0.4$. This value is inspired by several recent studies, e.g.\ $\sim 0.5$ in \citet{DRAINE14}, $0.5\pm0.1$ in \citet{CLARK16}, $0.4\pm0.2$ in \citet{CLARK19_2019MNRAS.489.5256C}, and $0.46^{+0.12}_{-0.06}$ in \citet{Chiang21}. For the very center of the galaxies ($R_\mathrm{g} = 0$), we assume an efficient dust growth, i.e.\ all refractory elements are completely depleted and gaseous elements (e.g.\ oxygen and nitrogen) are partially depleted, and adopted $\mathrm{D/M}=0.7$ from \citet{FELDMANN15}. In $R_\mathrm{e} \geq R_\mathrm{g} \geq 0$, we assume a smooth transition, that is:
\begin{equation}
    \mathrm{D/M} = \left\{\begin{array}{ll}
    0.7 - 0.3\times R_\mathrm{g}/R_\mathrm{e} & ,~R_\mathrm{g} \leq R_\mathrm{e} \\
    0.4 & ,~R_\mathrm{g} > R_\mathrm{e}
    \end{array}\right..
\end{equation}
With this toy model, we find $a=-0.74^{+0.06}_{-0.08}$ and $-0.47^{+0.04}_{-0.05}$ for \cotwo and \coone, respectively. These indices are steeper than our fiducial case, i.e.\ constant D/M, indicating that the \aco-to-\Sigmastar relation observed in \autoref{sec:aCO-SigmaMstar} is not caused by the variation in dust properties. Although we do expect an internal variation of D/M, the variation in the galaxy center predicted in simulations is more gentle than our toy model \citep{Romano22,Choban22}. Thus the calculation with this toy model should be interpreted as an extreme case. The constant D/M case and the toy model case should sandwich the actual indices.

\bibliographystyle{yahapj}
\bibliography{references}

%\listofchanges
\end{document}